\documentclass[OptBiber]{jtcam_preprint}

\usepackage{amsmath,mathtools}
\usepackage{graphicx}
\usepackage{longtable}
\usepackage{bm}
\usepackage{verbatim}

\makeatletter
\providecommand{\citet}{\@ifnextchar[{\citet@opt}{\citet@noopt}}
\def\citet@opt[#1]#2{\textcite[#1]{#2}}
\def\citet@noopt#1{\textcite{#1}}

\providecommand{\citep}{\@ifnextchar[{\citep@opt}{\citep@noopt}}
\def\citep@opt[#1]#2{\parencite[#1]{#2}}
\def\citep@noopt#1{\parencite{#1}}
\makeatother

\title{Propagation of weak layer failure in snow slab avalanche release: analytical solutions for a compliant interface with finite softening}

\runningtitle{Propagation of weak layer failure in snow slab avalanche release}

\author[1,2,3]{Johan Gaume}
\author[1,2,3]{Francis Meloche}
\author[4]{Ingrid Reiweger}
\author[5]{Pascal Hagenmuller}

\runningauthor{J. Gaume et al.}

\affil[1]{Institute for Geotechnical Engineering, ETH Z\"urich, Z\"urich, Switzerland}
\affil[2]{WSL Institute for Snow and Avalanche Research SLF, Davos, Switzerland}
\affil[3]{Climate Change, Extremes and Natural Hazards in Alpine Regions Research Center (CERC), Davos, Switzerland}
\affil[4]{Institute of Mountain Risk Engineering, Department of Landscape, Water and Infrastructure, BOKU University of Natural Resources and Life Sciences, 1180 Vienna, Austria}
\affil[5]{Universit\'e Grenoble Alpes, Universit\'e de Toulouse, M\'et\'eo-France, CNRS, CNRM, Centre d'Etudes de la Neige, Grenoble, France}

\keywords{snow slab avalanche, weak layer, shear failure, softening, anticrack, fracture process zone, material point method}

\begin{document}

\begin{abstract}
Snow slab avalanches are among the most dangerous natural hazards in mountain regions, causing most avalanche fatalities among recreationists and threatening settlements and infrastructure during large events. Recent numerical, field, and laboratory studies have renewed interest in shear-failure interpretations of avalanche release, particularly in relation to dynamic crack propagation and supershear fracture. Yet existing analytical shear models generally either assume a perfectly brittle transition from peak strength to residual friction, thereby neglecting post-peak dissipation, or neglect the pre-peak elastic response of the weak layer, which controls stress redistribution and critical crack length.\\

Here, we derive an analytical solution for shear failure propagation in a weak layer buried beneath an elastic snow slab, explicitly accounting for weak-layer elasticity and finite post-peak softening. Building on the brittle weak-spot model of \citet{Gaume2013a}, we consider a slab -- weak layer system composed of a fully softened residual zone, a linear fracture process zone, and an intact elastic region. In the brittle limit, the model recovers the classical weak-spot length \(a_{c0}\). For finite softening, it distinguishes between the fully softened crack length \(a_c\) and the total affected length \(b_c\), which also includes the process zone. This distinction is essential when comparing analytical predictions with numerical models or experiments.
The formulation links weak-spot models directly to sharp-crack fracture-energy descriptions, because fracture-energy effects enter through the finite-softening interface law rather than through an imposed propagation criterion. The exact solution can be written compactly as \(a_c=a_{c0}\sqrt{1+C_a\delta/u_p}\), where \(u_p\) is the displacement at peak strength and \(C_a\) depends on the peak, residual, and gravitational shear stresses. In the small-softening limit, the process-zone length \(\omega=b_c-a_c\) grows linearly with the softening displacement. Depth-averaged Material Point Method simulations using the same interface law confirm the analytical displacement and stress profiles and reproduce the predicted evolution of \(a_c\), \(b_c\), and \(\omega\). Laboratory shear data suggest that the softening displacement is small compared with \(u_p\), supporting the relevance of the brittle and small-softening limits.\\

We then extend the same compliant-softening framework to collapse-driven anticrack propagation. A simplified Timoshenko anticrack analogue shows that slab bending and transverse shear deformation control normal stress amplification at the weak-layer front; with finite compressive softening, this introduces a bending-controlled length scale and an approximately fourth-root dependence of anticrack length on softening displacement, supported by three-dimensional MPM simulations on flat terrain. Finally, we develop a mixed-mode Timoshenko anticrack formulation coupling weak-layer compression, slope-parallel shear, and slab rotation. A compact analytical model gives transparent sharp-front predictions, while a fully coupled finite-softening model resolves axial displacement, transverse deflection, rotation, normal support, and tangential support simultaneously. With realistic elastic properties and mixed-mode failure-envelope parameters, the resulting models reproduce the observed dependence of critical cut length on slope angle.
\end{abstract}

\section{Introduction}

Snow slab avalanche release is commonly described as a fracture process occurring in a weak snow layer buried beneath a cohesive slab. Although many modern theories emphasize weak-layer collapse and anticrack propagation \citep{Heierli2008}, the first analytical framework for snow slab avalanche release was fundamentally shear-based. Inspired by \citet{Palmer1973}, \citet{McClung1979} formulated the onset of crack propagation as the unstable extension of a shear band in a strain-softening weak layer, introducing the key idea that post-peak weakening provides the mechanical driving mechanism for release. This approach established a direct link between weak-layer strength, softening, slab load, and critical crack length. However, because the pre-peak elastic compliance of the weak layer was neglected, the critical length tends to zero in the limit of vanishing softening distance, or fracture energy, which is physically unrealistic.

Related shear-based models were later developed by \citet{Chiaia2008} and \citet{Gaume2013a}, the latter also including residual friction. In contrast to McClung's formulation, these models accounted for the elastic response of the weak layer before peak failure, but treated failure as an abrupt transition from peak strength to a weakened or residual state, without explicitly resolving a finite post-peak softening zone. Thus, existing shear-based analytical formulations either include finite softening without weak-layer pre-peak compliance, or include pre-peak compliance without finite softening. More importantly, because they remain fundamentally shear-based, they cannot explain key field observations such as remote triggering, whumpfs, and crack propagation on flat or low-angle terrain.

The latter limitation motivated the anticrack theory of \citet{Heierli2008}, in which weak-layer collapse induces slab bending and drives fracture with the closure of the crack faces. This anticrack framework successfully explained several observations that were difficult to reconcile with shear-only models and strongly influenced subsequent analytical and numerical work. The anticrack concept was later incorporated into numerous numerical models. In particular, the finite-element models of \citet{Mahajan2010,Volmer2017} and the discrete-element simulations of \citet{Gaume2015,Gaume2017,bobillier2024numerical} highlighted the importance of weak-layer collapse, slab bending, and mixed-mode fracture processes in governing both the onset and the subsequent dynamics of crack propagation. These studies progressively shifted much of the theoretical interpretation of slab avalanche release from a purely shear-driven instability toward a collapse-driven anticrack mechanism. More recently, efficient closed-form analytical models were introduced by \citet{rosendahl2020modeling1,rosendahl2020modeling2}. These models are based on a Timoshenko beam resting on a Winkler-type foundation representing the weak layer, with mode-I and mode-II fracture-energy contributions. They capture not only failure initiation but also the onset of crack propagation, including the effects of layering in the overlying slab \citep{weissgraeber2023closed}, overcoming some limitations of previous effective-stiffness models \citep{Monti2016}. However, these formulations treat fracture in a sharp-crack sense: dissipation is introduced through fracture-energy terms, while the post-peak degradation zone is not explicitly resolved through a local softening law with a non-zero softening displacement.

Recent work has shown that shear-based mechanisms remain essential, especially on steep slopes and at large propagation distances. Numerical simulations by \citet{trottet_np_2021} demonstrated a transition from sub-Rayleigh anticrack propagation to supershear crack propagation on slopes, indicating that once dynamic propagation is established, weak-layer collapse and slab bending may become secondary, with propagation then primarily driven by shear failure in the weak layer. This interpretation was further explored by \citet{guillet2023,meloche2025modeling}, who used a depth-averaged Material Point Method (MPM) with a weak-layer shear-softening interface to study the onset and dynamics of crack propagation, crack arrest, and avalanche release size at slope scale. In parallel, \citet{siron2023theoretical} provided a theoretical framework for dynamic crack propagation regimes, extending the unbounded theory of \citet{Heierli2008_} while explicitly incorporating weak-layer properties.

These recent developments bring renewed relevance to \citet{McClung1979}'s original shear concept. They suggest that although weak-layer collapse is crucial for crack nucleation and anticrack propagation on flat or low-angle terrain and during the early stages of crack propagation on slopes, the dynamics of large slab avalanches on steep slopes may be governed to first order by shear failure and softening in the weak layer. The objective of the present paper is therefore twofold. First, we revisit shear-failure propagation theory within a modern analytical framework that extends \citet{McClung1979}'s model to a situation in which the elasticity of both the slab and the weak layer is considered. By explicitly accounting for weak-layer compliance and post-peak softening over a non-zero displacement interval, we derive a comprehensive analytical model that connects weak-layer and slab mechanical properties to the onset of crack propagation. The resulting formulation provides a mechanics-based criterion for evaluating crack propagation propensity in a slab--weak-layer system, while accounting for weak-layer elasticity, strength, residual friction, and fracture energy through the constitutive softening law itself. Second, we extend the same compliant-softening framework to collapse-driven anticrack propagation. We derive a simplified Timoshenko anticrack analogue, develop both reduced analytical and fully coupled mixed-mode formulations coupling weak-layer compression and shear, and compare the resulting predictions with experimental critical cut lengths.

\section{Propagation of shear failure}
\label{sec:shear_growth}

We first note that, for clarity, we use \(a\) to denote the length of the region in which the weak layer has reached its residual strength, and \(a_c\) to denote the critical value of this length at the onset of crack propagation in a slab--weak-layer system, regardless of the assumed failure mode of the weak layer. Strictly speaking, in the case of weak-layer shear failure, and in light of recent numerical, theoretical, and experimental work, this length would more appropriately be referred to as the \emph{supercritical crack length} \(a_{sc}\), following \citet{trottet_np_2021} and \citet{meloche2025modeling}. However, to avoid introducing multiple notations, we deliberately retain the simpler notation \(a_c\) throughout this paper. In Section~\ref{sec:anticrack_growth}, where we extend the analysis to the onset of crack propagation under mixed-mode anticrack conditions, we continue to use the same notation for consistency.

\subsection{Shear failure with finite softening}
\label{sec:shear_softening}

\subsubsection{Mechanical framework}

We consider a one-dimensional slab--weak-layer system following the framework of \citet{Gaume2013a}, extended to include a finite post-peak softening displacement (Figure~\ref{fig:geometry}). The snow slab is modeled as a linear elastic layer of thickness \(h\), and deformation is assumed to occur only in the slope-parallel direction \(x\). We use a Cartesian coordinate system in which \(x\) is oriented slope-parallel and \(z\) is positive upward, with \(z=0\) at the free surface of the slab and \(z=-h\) at the slab--weak-layer interface. The gravitational loading is represented by a uniform shear stress \(\tau_g\), while the weak layer is represented as a shear interface transmitting a stress \(\tau(x)\). Stress redistribution along the weak layer is therefore governed by the axial elasticity of the overlying slab.

The slope-parallel equilibrium of a slab element can be written in integrated form as
\begin{equation}
\frac{\partial}{\partial x}
\left(
\int_{-h}^{0}\sigma_{xx}\,dz
\right)
-\tau(x)
=
-\tau_g ,
\end{equation}
where \(\sigma_{xx}\) is the slope-parallel normal stress in the slab. Under the one-dimensional elastic approximation, the depth-integrated (\(z\)-direction) axial force is related to the weak-layer displacement \(u(x)\) through
\[
\int_{-h}^{0}\sigma_{xx}\,dz \simeq E'h\,u'(x),
\]
where \(E'\) is the plane-strain elastic modulus. The governing equation is therefore
\begin{equation}
E'h\,u''(x)-\tau(x)=-\tau_g .
\label{eq:gov}
\end{equation}
Here and in the following, primes denote derivatives with respect to the slope-parallel coordinate \(x\), i.e. \(u'=du/dx\) and \(u''=d^2u/dx^2\).

\subsubsection{Weak-layer interface law}

\begin{figure*}[t]
\centerline{\includegraphics[width=0.9\textwidth, trim={0pt 360pt 0pt 0pt}, clip]{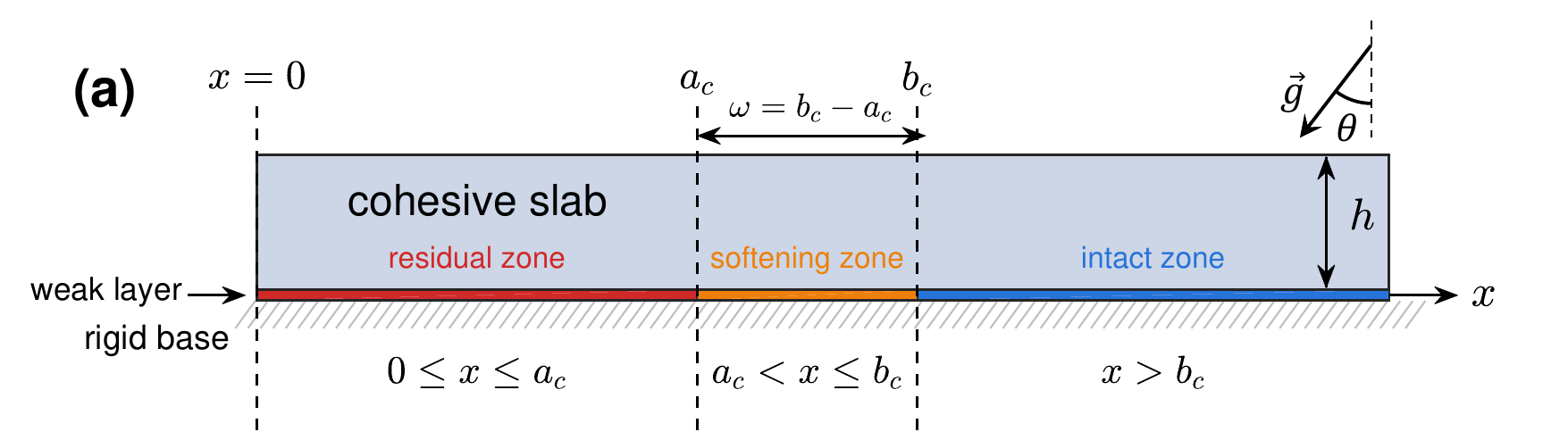}}
\centerline{\includegraphics[width=0.9\textwidth, trim={0pt 50pt 0pt 300pt}, clip]{figure1_new.pdf}}
\caption{(a) Slab--weak layer system geometry and (b) weak-layer constitutive model.}
\label{fig:geometry}
\end{figure*}

The weak layer is described by a local shear stress--displacement law with finite post-peak softening (Figure~\ref{fig:geometry}b). The response is linear elastic up to a peak shear stress \(\tau_p\), reached at displacement \(u_p\). It then softens linearly until the residual shear stress \(\tau_r\) is reached at displacement \(u_r\). The softening displacement is
\begin{equation}
\delta=u_r-u_p\geq0 .
\end{equation}
The interface law reads
\begin{equation}
\tau(u)=
\begin{cases}
\dfrac{\tau_p}{u_p}\,u, 
& 0\le u\le u_p,\\[2ex]
\tau_p-\dfrac{\tau_p-\tau_r}{\delta}(u-u_p), 
& u_p\le u\le u_r,\\[2ex]
\tau_r, 
& u\ge u_r .
\end{cases}
\label{eq:tau_u}
\end{equation}
This law provides a minimal representation of weak-layer weakening.

\subsubsection{Three-zone structure}

We consider the quasi-static critical configuration at the onset of crack propagation. In this configuration, the weak layer interface comprises a residual zone, a finite softening zone, and an intact zone, as illustrated in Figure~\ref{fig:geometry}a.

By geometrical symmetry of the idealized centered weak-spot configuration, we consider only the half-domain \(x\ge 0\). The three corresponding regions are described below. Closest to \(x=0\), the weak layer is fully softened and carries only the residual shear stress \(\tau_r\). This residual zone occupies
\[
0\le x\le a_c .
\]
Ahead of it lies a softening, or fracture process zone, in which the displacement decreases from \(u_r\) to \(u_p\) and the stress follows the softening branch of Eq.~\eqref{eq:tau_u}. This zone occupies
\[
a_c\le x\le b_c .
\]
Finally, for \(x\ge b_c\), the weak layer remains intact and follows the elastic branch.

The softening zone length is denoted by
\begin{equation}
\omega=b_c-a_c .
\end{equation}
Thus, \(a_c\) is the fully residual half-length, whereas \(b_c\) is the total affected half-length, including both the residual and softening zones. Symmetry imposes
\begin{equation}
u'(0)=0 .
\label{eq:sym}
\end{equation}
The symmetry assumption makes the problem equivalent to one half of a centered weak-spot configuration, rather than to the open-ended PST geometry commonly used in the field. This simplification is appropriate for the present shear model, but it would no longer hold when weak-layer collapse and the associated slab bending are explicitly considered, since bending breaks the left--right symmetry of the open-ended geometry.

\subsection{Exact solution and limiting behaviour}
\label{sec:shear_exact_limiting}

\subsubsection{Characteristic lengths}

The solution involves two characteristic lengths. The first is the elastic characteristic length of the slab--weak-layer system, which governs stress redistribution within the slab and the transfer of load to the weak layer,
\begin{equation}
\Lambda=
\sqrt{\frac{E'h\,u_p}{\tau_p}}.
\label{eq:Lambda}
\end{equation}
This is the same characteristic length appearing in the weak-spot model of \citet{Gaume2013a}. The second is the length associated with finite softening, or fracture process zone length,
\begin{equation}
\ell=
\sqrt{\frac{E'h\,\delta}{\tau_p-\tau_r}} .
\label{eq:ell}
\end{equation}

\subsubsection{Closed-form solution}

\paragraph{Critical lengths}

Solving the governing Eq.~\eqref{eq:gov} in the residual, softening, and intact regions gives three local second-order solutions, and therefore six integration constants. These constants are fixed by continuity of displacement and of the slab axial force \(E'h\,u'\) at the two internal boundaries \(x=a_c\) and \(x=b_c\), together with the symmetry condition \(u'(0)=0\) and the boundedness condition as \(x\to\infty\). Because the slab is linear elastic, continuity of axial force is equivalent here to continuity of \(u'\). The two front positions are then determined by the branch-transition conditions: at \(x=a_c\), the weak layer reaches the residual branch, \(u(a_c)=u_r\), or equivalently \(\tau(a_c)=\tau_r\); at \(x=b_c\), it reaches the peak point, \(u(b_c)=u_p\), or equivalently \(\tau(b_c)=\tau_p\). This gives a scalar equation (see complete derivation in Appendix~\ref{app:matching}) for the dimensionless softening-zone length
\begin{equation}
\alpha=\frac{b_c-a_c}{\ell}.
\label{eq:alpha_def}
\end{equation}

For the finite three-zone solution considered here, we require
\[
\tau_r<\tau_g<\tau_p .
\]
The condition \(\tau_g<\tau_p\) ensures that the far-field weak layer remains on the intact elastic branch, whereas \(\tau_g>\tau_r\) provides the net driving stress required to sustain a finite residual zone.

The resulting matching equation is
\begin{equation}
(\tau_g-\tau_r)
+
(\tau_p-\tau_g)\cos\alpha
=
(\tau_p-\tau_g)\frac{\Lambda}{\ell}\sin\alpha .
\label{eq:alpha_eq}
\end{equation}
Equivalently,
\begin{equation}
\alpha=
\arccos\left(
-\frac{\tau_g-\tau_r}
{(\tau_p-\tau_g)\sqrt{1+(\Lambda/\ell)^2}}
\right)
-\arctan\left(\frac{\Lambda}{\ell}\right).
\label{eq:alpha_closed}
\end{equation}
The fully residual half-length is then
\begin{equation}
a_c=
\ell\,
\frac{
\dfrac{\tau_p-\tau_g}{\tau_g-\tau_r}
+\cos\alpha
}{
\sin\alpha
},
\label{eq:ac_closed}
\end{equation}
and the total affected half-length is
\begin{equation}
b_c=a_c+\ell\alpha .
\label{eq:bc_closed}
\end{equation}

\begin{figure*}[t]
\centerline{\includegraphics[width=0.7\textwidth, trim={40pt 270pt 40pt 270pt}, clip]{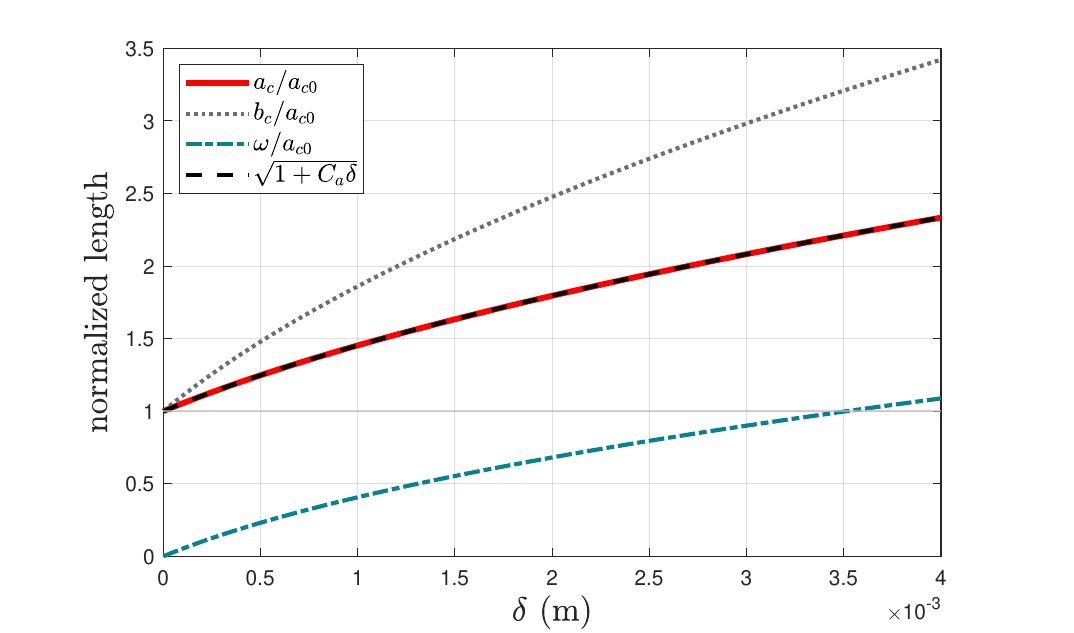}}
\caption{Normalized characteristic lengths as a function of the softening displacement \(\delta\). The critical crack length \(a_c\), the total affected length \(b_c\), and the softening-zone length \(\omega=b_c-a_c\) are normalized by the brittle critical length \(a_{c0}\). The dashed curve shows the exact compact square-root expression for \(a_c/a_{c0}\), given by \(\sqrt{1+C_a\delta/u_p}\). The results illustrate that finite softening modifies the fully softened crack length and introduces a finite process zone, so that the total affected length \(b_c\) exceeds the critical crack length \(a_c\). The mechanical parameters used are \(E'=1.0~\mathrm{MPa}\), \(h=0.60~\mathrm{m}\), \(u_p=1.0~\mathrm{mm}\), \(\tau_p=2.5~\mathrm{kPa}\), \(\tau_r=0.5~\mathrm{kPa}\), and \(\tau_g=1.0~\mathrm{kPa}\).}
\label{fig:shear_lengths}
\end{figure*}

The expression for \(a_c\) can be further simplified (see also Appendix~\ref{app:small_delta}). Introducing
\[
T=\tau_p-\tau_g,
\qquad
R=\tau_g-\tau_r,
\qquad
s=\frac{\ell}{\Lambda},
\]
the exact matching equations imply
\begin{equation}
\left(\frac{a_c}{\ell}\right)^2
=
\left(\frac{T}{R}\right)^2
\left(1+\frac{1}{s^2}\right)-1 .
\label{eq:ac_over_ell_exact}
\end{equation}
Using
\[
a_{c0}=\Lambda\frac{T}{R}
=
\Lambda
\frac{\tau_p-\tau_g}{\tau_g-\tau_r}
\]
and
\[
\frac{\ell^2}{\Lambda^2}
=
\frac{\tau_p}{\tau_p-\tau_r}\frac{\delta}{u_p},
\]
the fully residual crack length can therefore be written exactly as
\begin{equation}
a_c
=
a_{c0}
\sqrt{
1+C_a\frac{\delta}{u_p}
},
\label{eq:ac_sqrt_exact}
\end{equation}
with
\begin{equation}
C_a
=
\frac{
\tau_p(\tau_p-2\tau_g+\tau_r)
}{
(\tau_p-\tau_g)^2
}.
\label{eq:Ca_main}
\end{equation}
Thus, the square-root dependence is an exact consequence of the three-zone analytical solution. By contrast, the total affected length \(b_c\) remains controlled by the exact process-zone contribution \(\omega=\ell\alpha\), where \(\alpha\) is given by Eq.~\eqref{eq:alpha_closed}.

Finally, the fracture process zone length is defined as
\[
\omega = \ell \alpha = b_c-a_c .
\]

Figure~\ref{fig:shear_lengths} shows the evolution of the critical crack length \(a_c\),
the total affected length \(b_c\), and the softening length \(\omega\), all
normalized by the brittle critical length \(a_{c0}\), as functions of the softening
distance \(\delta\).

\paragraph{Tangential displacement and shear stress profiles}

\begin{figure*}[t]
\centerline{\includegraphics[width=\textwidth, trim={0pt 40pt 0pt 40pt}, clip]{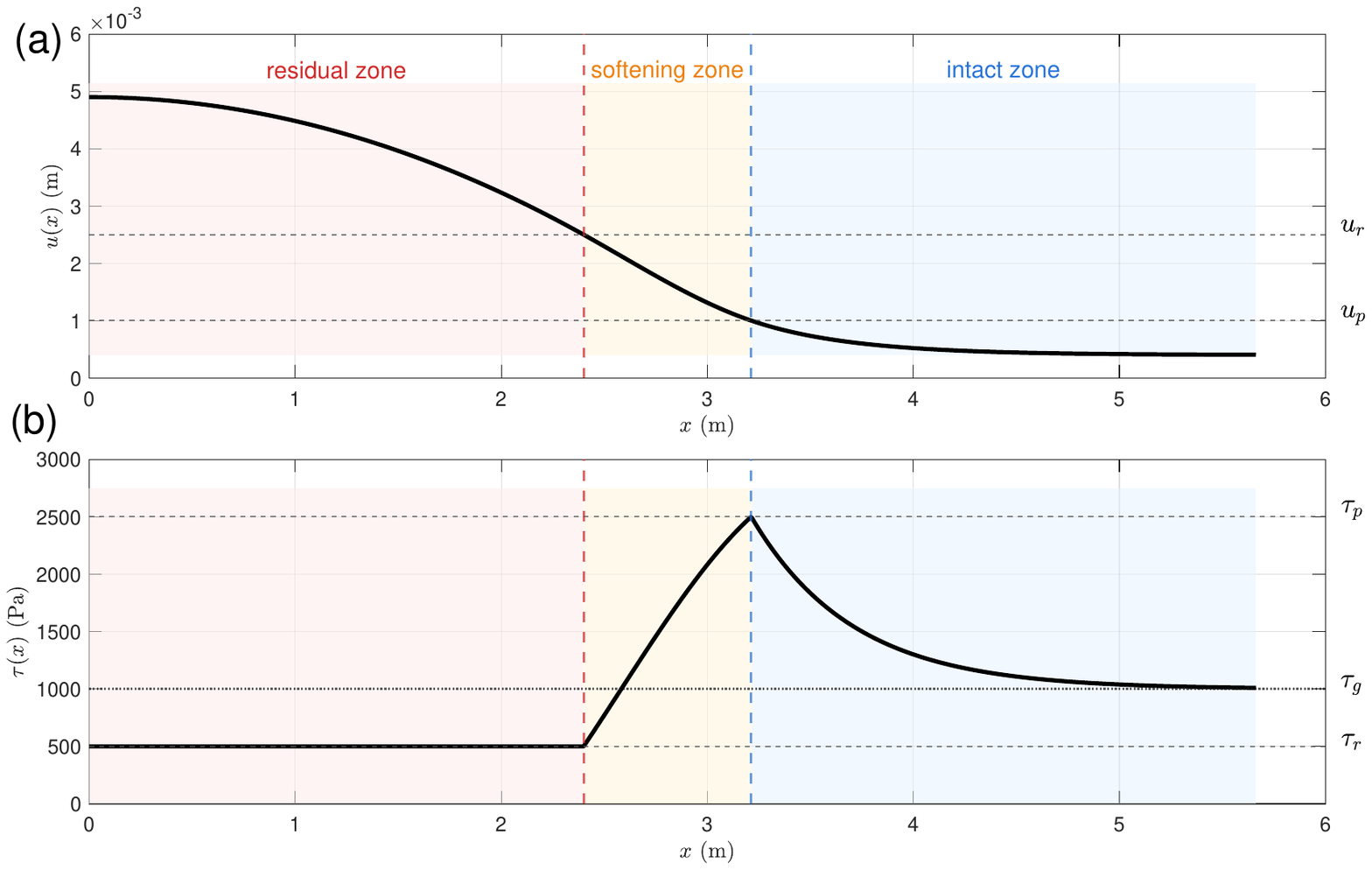}}
\caption{Solution for (a) the tangential displacement \(u(x)\) and (b) shear stress \(\tau(x)\) in the three regions for the same set of mechanical parameters as in Figure~\ref{fig:shear_lengths}.}
\label{fig:shear_profiles}
\end{figure*}

The corresponding displacement and stress fields at the critical state are obtained as follows. In the residual zone, \(0\le x\le a_c\), the weak layer carries the constant residual shear stress \(\tau_I=\tau_r\), and the displacement is
\begin{equation}
u_I(x)=u_r+\frac{\tau_r-\tau_g}{2E'h}\left(x^2-a_c^2\right).
\label{eq:uI_main}
\end{equation}
This expression directly satisfies \(u_I(a_c)=u_r\).

In the softening zone, \(a_c\le x\le b_c\), the displacement takes the form
\begin{equation}
u_{II}(x)=u_c+A\cos\left(\frac{x-a_c}{\ell}\right)+B\sin\left(\frac{x-a_c}{\ell}\right),
\label{eq:uII_main}
\end{equation}
with
\[
u_c=u_p+\delta\frac{\tau_p-\tau_g}{\tau_p-\tau_r},
\qquad
A=\delta\frac{\tau_g-\tau_r}{\tau_p-\tau_r},
\qquad
B=-A\frac{a_c}{\ell}.
\]

Finally, in the intact zone, \(x\ge b_c\), the bounded displacement field is
\begin{equation}
u_{III}(x)=\frac{\tau_g}{\tau_p}u_p+u_p\frac{\tau_p-\tau_g}{\tau_p}e^{-(x-b_c)/\Lambda}.
\label{eq:uIII_main}
\end{equation}

The corresponding weak-layer shear stresses in the three regions follow directly by substituting \(u_I\), \(u_{II}\), and \(u_{III}\) into the interface law in Eq.~\eqref{eq:tau_u}. The complete displacement and stress profiles are illustrated in Figure~\ref{fig:shear_profiles}. The detailed derivation of Eqs.~\eqref{eq:alpha_eq}--\eqref{eq:bc_closed} is given in Appendix~\ref{app:matching}.

\subsubsection{Brittle limit}

In the brittle limit \(\delta\to 0\), the fracture process zone length \(\ell\) tends to zero while \(\Lambda\) remains finite. Hence \(\Lambda/\ell\to\infty\). Equation~\eqref{eq:alpha_eq} then requires \(\alpha\to 0\), so that the process zone length \(\omega=\ell\alpha\) also vanishes. The residual-zone length and the total affected length therefore become identical.

The brittle limit follows directly from Eq.~\eqref{eq:ac_sqrt_exact}. As \(\delta\to0\), the fully residual crack length satisfies
\begin{equation}
a_c \to a_{c0}
=
\Lambda
\frac{\tau_p-\tau_g}{\tau_g-\tau_r}.
\label{eq:ac0}
\end{equation}
Moreover, Eq.~\eqref{eq:alpha_eq} gives \(\alpha=O(\ell/\Lambda)\), so that the process-zone length \(\omega=\ell\alpha\) vanishes and
\[
b_c=a_c+\omega\to a_{c0}.
\]
Thus, the present fracture-process-zone solution recovers the classical brittle weak-spot result in the limit of vanishing softening displacement \citep{Gaume2013a}.

\subsubsection{Small-softening behaviour of the process zone \texorpdfstring{\(\delta \rightarrow 0\)}{delta -> 0}}

Although the fully residual crack length \(a_c\) admits the exact compact form
given by Eq.~\eqref{eq:ac_sqrt_exact}, the process-zone length
\(\omega=b_c-a_c=\ell\alpha\) still depends on the matching angle \(\alpha\).
It is therefore useful to examine its small-softening limit.

For \(\delta\to0\), one has \(\ell/\Lambda\to0\) and \(\alpha\to0\). Expanding
Eq.~\eqref{eq:alpha_eq} gives
\begin{equation}
\alpha
\sim
\frac{\tau_p-\tau_r}{\tau_p-\tau_g}
\frac{\ell}{\Lambda}.
\end{equation}
Since \(\omega=\ell\alpha\), this yields
\begin{equation}
\omega
\simeq
\Lambda\,\frac{\tau_p}{\tau_p-\tau_g}\frac{\delta}{u_p}.
\label{eq:omega_small_delta}
\end{equation}
Thus, although the intrinsic softening length \(\ell\) scales as
\(\delta^{1/2}\), the actual process-zone length
\(\omega=b_c-a_c\) grows linearly with the softening displacement in the
small-\(\delta\) regime.

The exact expression for \(a_c\), Eq.~\eqref{eq:ac_sqrt_exact}, can also be
linearized for small \(\delta/u_p\). Using
\(\sqrt{1+x}=1+x/2+O(x^2)\), one obtains
\begin{equation}
a_c
\simeq
a_{c0}
\left(
1+\frac{C_a}{2}\frac{\delta}{u_p}
\right),
\label{eq:ac_linear_small_delta}
\end{equation}
where
\begin{equation}
C_a
=
\frac{
\tau_p(\tau_p-2\tau_g+\tau_r)
}{
(\tau_p-\tau_g)^2
}.
\end{equation}
The coefficient \(C_a\) changes sign when
\(\tau_g=(\tau_p+\tau_r)/2\). Accordingly, finite softening increases the
fully residual critical length if \(\tau_g<(\tau_p+\tau_r)/2\), but decreases
it if \(\tau_g>(\tau_p+\tau_r)/2\).

The total affected length follows exactly from
\begin{equation}
b_c=a_c+\omega,
\label{eq:bc_from_ac_omega}
\end{equation}
when \(a_c\) is given by Eq.~\eqref{eq:ac_sqrt_exact} and
\(\omega=\ell\alpha\) is evaluated using Eq.~\eqref{eq:alpha_closed}. In the
small-softening limit, combining Eqs.~\eqref{eq:omega_small_delta} and
\eqref{eq:ac_linear_small_delta} gives
\begin{equation}
b_c
\simeq
a_{c0}
\left(
1+\frac{C_b}{2}\frac{\delta}{u_p}
\right),
\label{eq:bc_linear_small_delta}
\end{equation}
where
\begin{equation}
C_b
=
\frac{
\tau_p(\tau_p-\tau_r)
}{
(\tau_p-\tau_g)^2
}.
\label{eq:Cb}
\end{equation}

Unlike \(C_a\), the coefficient \(C_b\) is positive for \(\tau_p>\tau_r\).
Therefore, the total affected length \(b_c\) always increases with softening
displacement in the small-\(\delta\) regime. This distinction is important when
comparing the analytical solution to numerical simulations. If the numerical
model measures the full damaged or affected region, including the process zone,
then \(b_c\), rather than \(a_c\), is the appropriate analytical quantity for
comparison.

\subsection{Verification with DAMPM}
\label{sec:dampm_verification}

Simulations based on the depth-averaged Material Point Method (DAMPM) were performed using the same shear softening law as that assumed in the analytical model. 
The main difference lies in the boundary condition at \(x=0\): instead of imposing symmetry in the numerical setup, we explicitly simulated a centered weak spot. In addition, the DAMPM simulations are not strictly quasi-static. Even for very slow cutting, a small slab acceleration develops within the residual zone. To account for this effect, we introduced a local inertial correction term \(\rho h\,\ddot u_{\mathrm{res}}\) on the right-hand side of Eq.~\eqref{eq:gov} and only in the residual region, with \(\ddot u_{\mathrm{res}}\) determined in DAMPM. Hence, the correction modifies only the residual-zone curvature and the value of \(u(0)\), while leaving the matching at \(x=a_c\), and therefore the analytical values of \(a_c\) and \(b_c\), unchanged.

\begin{figure*}[t]
\centerline{\includegraphics[width=1.0\textwidth, trim={40pt 120pt 80pt 120pt}, clip]{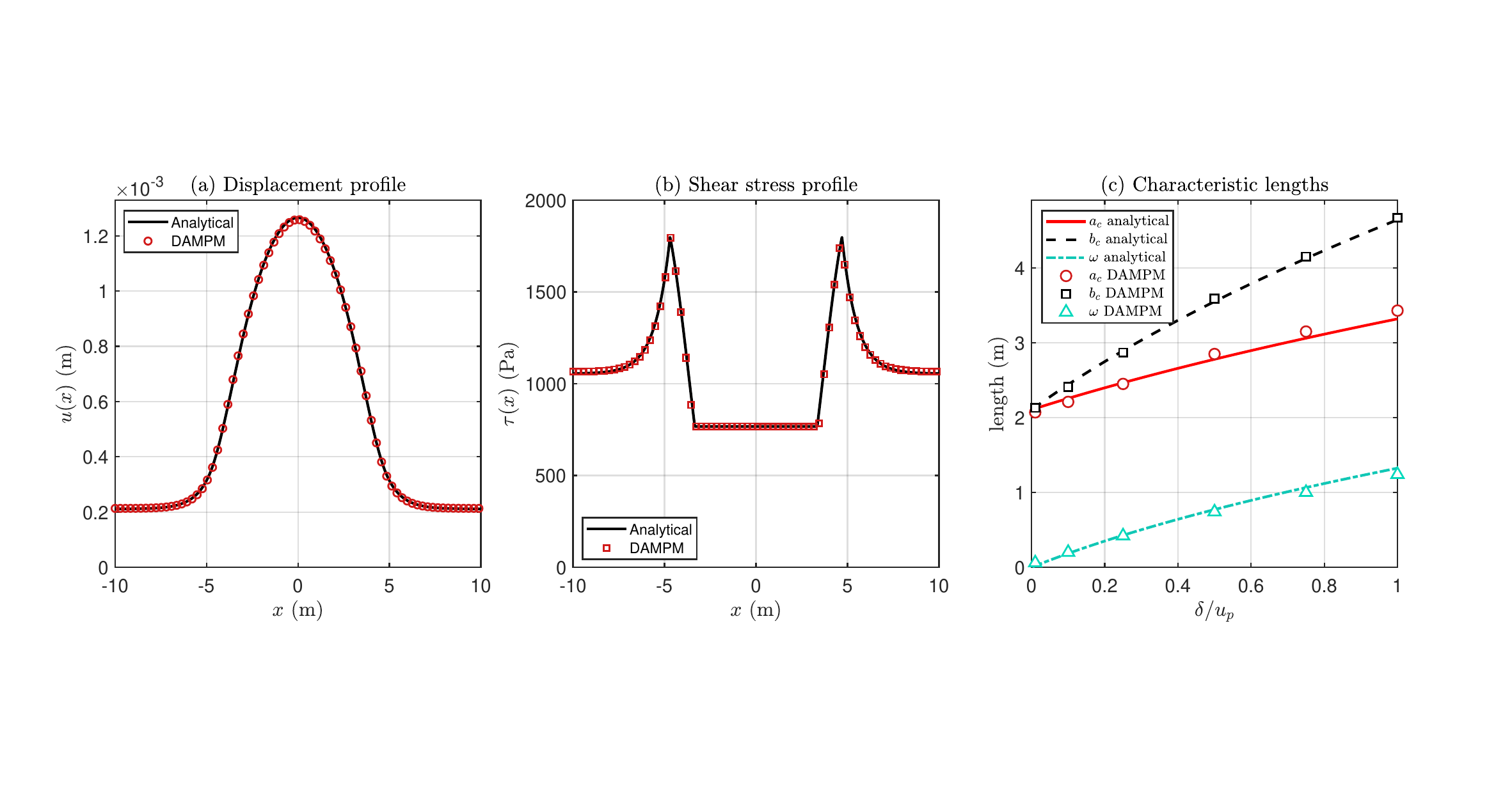}}
\caption{Comparison between the analytical slab--weak-layer softening solution and DAMPM results. (a) Tangential displacement profile \(u(x)\) for \(\delta/u_p=1\). (b) Corresponding shear stress profile \(\tau(x)\). (c) Critical crack length \(a_c\), total affected length \(b_c\), and fracture-process-zone length \(\omega=b_c-a_c\) as functions of the normalized softening displacement \(\delta/u_p\). Analytical predictions are shown by black lines and DAMPM results by red symbols. In panels (a) and (b), the analytical solution includes a local residual-zone dynamic correction to improve comparison with the non-quasi-static DAMPM profiles, while the critical lengths in panel (c) correspond to the original quasi-static analytical solution. For readability, the DAMPM profiles in panels (a) and (b) are shown using subsampled markers; these symbols do not represent the actual numerical grid points. The simulations were performed with a spatial resolution of \(2~\mathrm{cm}\), for which plotting every point would make the markers overlap and obscure the comparison.}
\label{fig:dampm_validation}
\end{figure*}

Overall, the agreement between analytical and numerical results is very good (Fig.~\ref{fig:dampm_validation}). In particular, the tangential displacement and shear stress profiles compare well (Figs.~\ref{fig:dampm_validation}a and \ref{fig:dampm_validation}b), and the analytical model accurately reproduces the dependence of the critical crack length \(a_c\), the total affected length \(b_c\), and the fracture-process-zone length \(\omega=b_c-a_c\) on the softening displacement (Fig.~\ref{fig:dampm_validation}c).

\subsection{Physical interpretation}

The present solution provides a bridge between two limiting descriptions of weak-layer shear failure. In the brittle limit \(\delta\to 0\), the weak layer drops abruptly from peak to residual strength and the formulation reduces to the brittle weak-spot criterion of \citet{Gaume2013a}. For finite \(\delta\), a softening zone develops ahead of the fully residual region, thereby introducing into the weak-spot framework the strain-softening picture originally emphasized by \citet{McClung1979}. In this sense, the model extends the brittle weak-spot approach to finite softening without the need to impose an additional fracture criterion externally: fracture-energy effects enter naturally through the constitutive law and the resulting emergence of a softening zone often referred to as fracture process zone in the literature. In the limiting case \(u_p=0\), the weak layer becomes non-compliant prior to peak failure, so that the elastic characteristic length associated with pre-peak stress redistribution disappears. The problem is then governed entirely by post-peak softening and by the associated fracture-process-zone length. This limiting case should therefore not be viewed as recovering exactly the formulation of \citet{McClung1979}, but rather as recovering the same class of post-peak shear-propagation models. In this respect, the present framework is also close in spirit to the process-zone and energy-balance approaches of \citet{puzrin2005growth}, although the result is obtained here directly from an exact slab--weak-layer constitutive solution.

A central result is that two crack-length measures must be distinguished. The length \(a_c\) denotes only the portion of the weak layer that has already reached residual strength, whereas \(b_c\) denotes the total affected length, including both the residual region and the softening zone. These two quantities should therefore not be used interchangeably. This distinction is particularly important when comparing the analytical solution with numerical simulations or observations, because the measured crack length may correspond to the full damaged region rather than to the fully residual part alone. In the present model, \(b_c\) always increases with \(\delta\) in the small-softening regime, reflecting the finite width of the softening zone. By contrast, the exact square-root expression for \(a_c\) is governed by
\[
C_a
=
\frac{
\tau_p(\tau_p-2\tau_g+\tau_r)
}{
(\tau_p-\tau_g)^2
},
\]
which changes sign when \(\tau_g=(\tau_p+\tau_r)/2\). Accordingly, if
\(\tau_g<(\tau_p+\tau_r)/2\),
finite softening increases the fully residual critical length, whereas if
\(\tau_g>(\tau_p+\tau_r)/2\),
it decreases it.

This latter result may appear counterintuitive, since introducing softening also introduces a fracture-energy-like dissipation, and one might therefore expect the critical crack length to increase systematically. The solution shows that this intuition applies more directly to the total affected length \(b_c\), not necessarily to the fully residual length \(a_c\). Indeed, \(a_c\) measures only the part of the weak layer that has already reached residual strength, while the softening zone ahead of it already contributes to stress redistribution and weakening. As a result, when the gravitational shear stress is sufficiently close to the peak strength, the system may become unstable with a shorter fully residual core because part of the required weakening is already provided by the softening zone. This does not mean that fracture energy promotes failure in an energetic sense; rather, it means that the relevant measure of the destabilized region is no longer \(a_c\) alone. The physically more complete quantity is \(b_c=a_c+\ell\alpha\), which accounts for the full region involved in the failure process.

The model remains deliberately idealized: it is one-dimensional, static, and based on a local shear stress--displacement law. Nevertheless, it provides a compact analytical benchmark for numerical models and clarifies how finite softening modifies the critical conditions for shear-failure propagation, while also showing that the interpretation of ``critical crack length'' depends on whether one refers to the fully residual region or to the total affected zone.

\subsection{Experimental constraints on the softening displacement}
\label{sec:softening_experimental_constraints}

Stress--deformation measurements of weak snow layers remain scarce. \citet{McClung1979} reported a shear stress--tangential displacement curve for a thin dry snow sample, showing pronounced post-peak softening and introducing the concept of a finite softening distance \(\delta\). However, the experiments published in \citet{McClung1979} and described in detail by \citet{McClung1977} were performed at very low strain rates, below \(10^{-4}\,\mathrm{s^{-1}}\), which is not representative of the nearly brittle response expected during artificial avalanche triggering or in classical snow-fracture tests. These experiments, and the conclusions drawn from them, are therefore likely more relevant to natural avalanche release, for which the ductile-to-brittle transition of snow may play an important role \citep{Puzrin2019}. More recently, \citet{schottner2025testing} presented stress--strain curves for depth-hoar weak layers that also exhibit pronounced post-peak softening with a finite and non-negligible softening length. These experiments were performed closer to the brittle regime than the low-rate tests of \citet{McClung1979,McClung1977}. However, the strain rates inferred from digital image correlation were still of the order of \(10^{-3}\,\mathrm{s^{-1}}\), so some ductile contribution to the measured softening response cannot be ruled out \citep{Puzrin2019}.

\begin{figure*}[t]
\centerline{\includegraphics[width=1.0\textwidth]{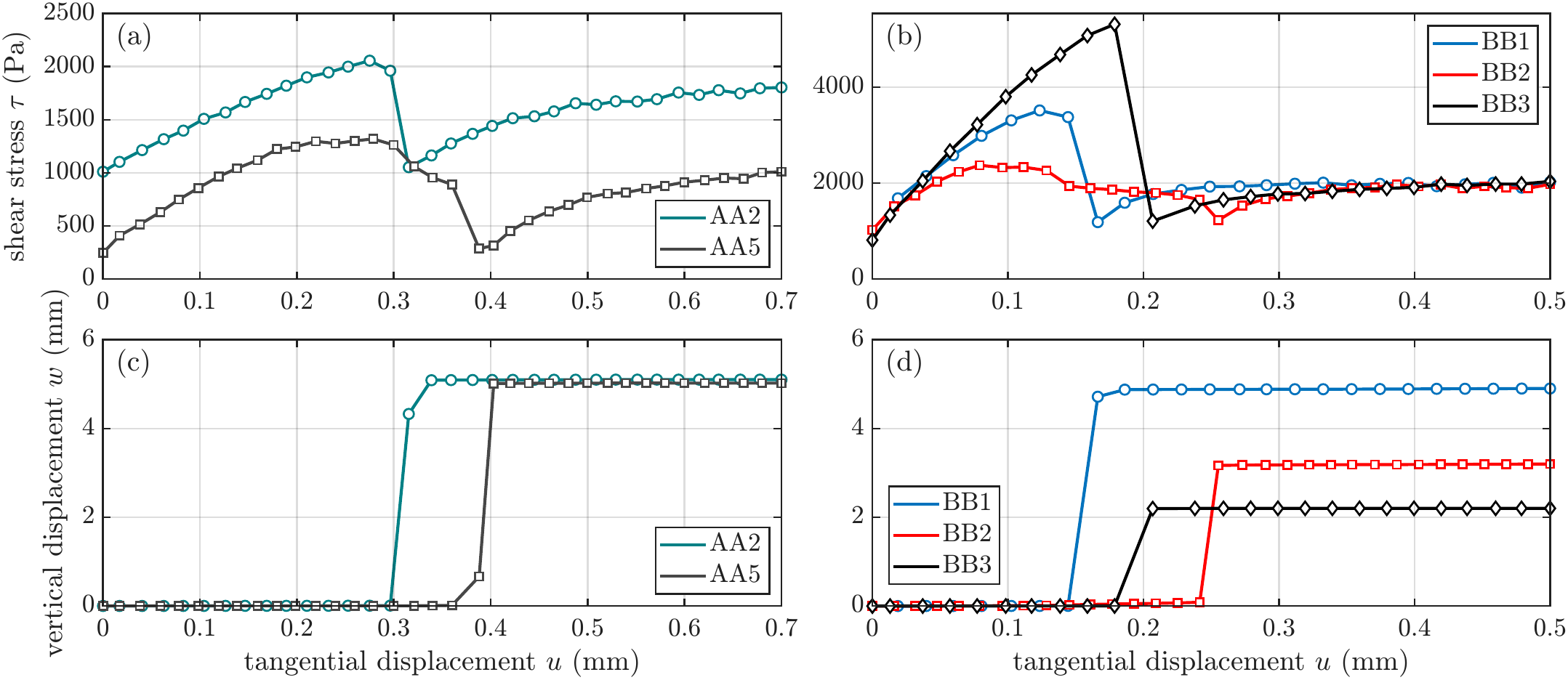}}
\caption{Displacement-controlled direct shear experiments on AA and BB snow samples from the thesis of Ingrid Reiweger \protect\citep{reiweger2011failure}. (a) Shear stress versus tangential displacement for the AA samples. (b) Shear stress versus tangential displacement for the BB samples. (c) Vertical displacement versus tangential displacement for the AA samples. (d) Vertical displacement versus tangential displacement for the BB samples. The abrupt post-peak stress drops indicate that the apparent softening distance is strongly limited by the acquisition frequency and should therefore be interpreted as an upper bound.}
\label{fig:reiweger_tests}
\end{figure*}

At higher loading rates, one may instead expect an even sharper, more brittle-like response. This interpretation is supported by the experimental results of \citet{Schweizer1998} and by displacement-controlled direct shear tests reported in \citet{reiweger2011failure}. The AA and BB samples were tested at strain rates ranging from $8.5\times10^{-3}$ to $1.3\times10^{-2}~\mathrm{s}^{-1}$ and exhibited a pronounced and abrupt post-peak stress drop, as illustrated in Fig.~\ref{fig:reiweger_tests}. In particular, the BB samples show a direct transition from peak to residual shear stress over an apparently very small tangential displacement. These observations suggest values of \(\delta\) that are significantly smaller than the peak displacement \(u_p\). Reiweger's experiments suggest a maximum ratio \(\delta/u_p \approx 0.1\). This value should, however, be regarded as an upper bound, since it is essentially limited by the acquisition frequency: no data point is available within the softening branch itself, and the measurements only capture a direct drop from peak to residual stress. In practice, the true value of \(\delta\) may therefore be smaller.

Within the present shear-based weak-layer interface framework, even the upper-bound value \(\delta/u_p \approx 0.1\) would lead to only a limited increase in the critical length, of the order of \(10\%\) at most for the parameters considered here. This suggests that, for supershear crack propagation driven primarily by weak-layer shear failure, the brittle weak-spot limit of the finite-softening interface formulation may be sufficient as a first approximation. Whether the same conclusion holds for collapse-driven anticrack nucleation is examined in Section~\ref{sec:anticrack_growth}.

Overall, these considerations suggest that the brittle critical length may provide a useful first-order approximation for shear-driven propagation in dry persistent weak layers under rapid loading. This conclusion should, however, be interpreted with care. The experimental constraints discussed above mainly concern dry, persistent weak layers loaded at high strain rates, whereas the effective softening distance may depend strongly on weak-layer type, liquid-water content, temperature, and loading rate. In particular, lower-rate experiments have reported softening distances of the same order as the elastic displacement \(u_p\), for which finite-softening corrections could become much more important. The present results should therefore not be generalized without caution to non-persistent weak layers, wet-snow instabilities, or slowly loaded natural release scenarios. In such cases, the brittle limit may no longer be sufficient and an explicit finite-softening description may be required.

\subsection{Connection to LEFM}

The present finite-softening weak-spot formulation can be interpreted as a cohesive-zone-type regularization of weak-layer shear failure. The weak-layer interface law resolves explicitly the transition from peak strength to residual strength over a finite displacement interval \(\delta\). This differs from a sharp-crack LEFM description, in which the finite softening zone is collapsed into an effective crack tip and the failure process is represented through an energy-release-rate condition. The first-order equivalence between the resolved process-zone solution and the effective sharp-crack description is therefore expected in the small-scale-yielding (SSY) limit, where the process-zone length is small compared with the elastic transfer length, \(\omega\ll\Lambda\).

For the linear interface law adopted here, it is useful to distinguish between three energetic contributions: the mechanical energy release rate of the slab--weak-layer system, the intrinsic bond-breaking energy of the weak layer, and the residual frictional work along the failed interface.

The intrinsic energy associated with breaking the cohesive bonds of the weak layer is the part of the stress--displacement work that degrades the interface from the intact state to the fully broken state. For the piecewise linear law used here, this intrinsic contribution can be written as
\begin{equation}
G_{IIc}^{\mathrm{intr}}=\frac12\tau_p u_p+\frac12\tau_p\delta .
\label{eq:Gintr}
\end{equation}
This quantity represents the work associated with the loss of cohesive resistance from the intact interface to the fully broken state. It does not include the subsequent residual sliding resistance of the failed weak layer.

The residual stress \(\tau_r\) instead represents the shear resistance carried by the already failed interface. The corresponding frictional work over the displacement interval relevant to the cohesive degradation is
\begin{equation}
W_{\mathrm{fric}}=\tau_r\left(u_p+\frac12\delta\right).
\label{eq:Wfric}
\end{equation}
This contribution should not be confused with intrinsic fracture energy. In particular, if the residual stress is interpreted as Coulomb friction, then \(\tau_r\) depends on normal load and therefore on slope angle and slab thickness. It is then load dependent and cannot be treated as an intrinsic material toughness.

In the finite-softening interface formulation used in the present paper, the post-peak softening contribution measured above the residual stress level is
\begin{equation}
G_{\mathrm{soft}}=\int_{u_p}^{u_r}\bigl(\tau(u)-\tau_r\bigr)\,du=\frac12(\tau_p-\tau_r)\delta .
\label{eq:Gsoft}
\end{equation}
This is the part of the post-peak work associated with strength degradation from \(\tau_p\) to \(\tau_r\). It gives
\begin{equation}
\delta=\frac{2G_{\mathrm{soft}}}{\tau_p-\tau_r},
\end{equation}
and the shear softening length can therefore be written as
\begin{equation}
\ell=\frac{\sqrt{2E'h\,G_{\mathrm{soft}}}}{\tau_p-\tau_r}.
\label{eq:ell_G}
\end{equation}
Thus, the finite post-peak softening zone is directly controlled by the cohesive degradation energy \(G_{\mathrm{soft}}\), whereas the residual stress controls the frictional resistance carried by the already failed interface.

Equivalently, one may combine the intrinsic bond-breaking contribution and the residual frictional contribution into an effective fracture energy for a sharp-crack LEFM-type formulation. In such a formulation, the process zone is not resolved explicitly. The energy balance therefore represents the replacement of the intact elastic interface by a residual interface in the crack wake. For the present shear law, the corresponding effective mode-II fracture energy is
\begin{equation}
G_{IIc}^{\mathrm{LEFM}}=G_{IIc}^{\mathrm{intr}}-W_{\mathrm{fric}}=\left(\frac12\tau_p-\tau_r\right)u_p+\frac12(\tau_p-\tau_r)\delta .
\label{eq:GLEFM}
\end{equation}
This expression is equivalent to
\begin{equation}
G_{IIc}^{\mathrm{LEFM}}=\int_{0}^{u_r}\bigl(\tau(u)-\tau_r\bigr)\,du .
\end{equation}
It is therefore an effective sharp-crack quantity, not necessarily an intrinsic material toughness. The decomposition into \(G_{IIc}^{\mathrm{intr}}\) and \(W_{\mathrm{fric}}\) is particularly useful when \(\tau_r\) is Coulomb-like, because it separates the intrinsic bond-breaking work from the load-dependent frictional work.

We now make the link with a LEFM energy balance explicit. Consider the corresponding sharp-crack problem in which the weak layer is fully failed over \(0\le x\le a\) and intact for \(x\ge a\). In the failed region, the interface carries only the residual stress \(\tau_r\), while in the intact region it remains linearly elastic with stiffness \(\tau_p/u_p\). The displacement fields are therefore the sharp-crack counterparts of the residual and intact solutions given above: in the failed region the solution is parabolic, as in Eq.~\eqref{eq:uI_main}, whereas in the intact region it decays exponentially over the elastic transfer length \(\Lambda\), as in Eq.~\eqref{eq:uIII_main}. For an arbitrary prescribed crack length \(a\), enforcing continuity of displacement and axial force at \(x=a\) gives
\begin{equation}
u_I(x,a)=u_0+\frac{\tau_r-\tau_g}{2E'h}x^2,\qquad 0\le x\le a,
\label{eq:uI_LEFM}
\end{equation}
and
\begin{equation}
u_{II}(x,a)=\frac{\tau_g}{\tau_p}u_p+C\,e^{-(x-a)/\Lambda},\qquad x\ge a,
\label{eq:uII_LEFM}
\end{equation}
with
\begin{equation}
C=\frac{a u_p(\tau_g-\tau_r)}{\Lambda\tau_p},\qquad u_0=\frac{u_p}{2\Lambda^2\tau_p}\left[2\Lambda^2\tau_g+2\Lambda a(\tau_g-\tau_r)+a^2(\tau_g-\tau_r)\right].
\label{eq:Cu0_LEFM}
\end{equation}

The total potential energy per unit width is the sum of slab strain energy, intact-interface elastic energy, gravitational work, and residual-interface work. Since the intact solution tends to a non-zero far-field displacement, this potential energy is understood up to an arbitrary \(a\)-independent reference energy. This constant does not affect the energy-release rate \(G(a)=-d\Pi/da\). We therefore write
\begin{equation}
\Pi(a)=\int_0^a\left[\frac12E'h\,u_I'^2-(\tau_g-\tau_r)u_I\right]dx+\int_a^\infty\left[\frac12E'h\,u_{II}'^2+\frac12\frac{\tau_p}{u_p}u_{II}^2-\tau_g u_{II}\right]dx .
\label{eq:Pi_LEFM}
\end{equation}
The corresponding energy-release rate is
\begin{equation}
G(a)=-\frac{d\Pi}{da}=\frac{u_p}{2\tau_p}\left[(\tau_g-\tau_r)^2\left(1+\frac{a}{\Lambda}\right)^2-\tau_r^2\right].
\label{eq:G_LEFM}
\end{equation}
The LEFM critical length \(a_{\mathrm{LEFM}}\) is then obtained from
\begin{equation}
G(a_{\mathrm{LEFM}})=G_{IIc}^{\mathrm{LEFM}} .
\label{eq:LEFM_balance}
\end{equation}
Solving Eq.~\eqref{eq:LEFM_balance} gives
\begin{equation}
a_{\mathrm{LEFM}}=\Lambda\left[\frac{1}{\tau_g-\tau_r}\sqrt{\frac{2\tau_p}{u_p}G_{IIc}^{\mathrm{LEFM}}+\tau_r^2}-1\right].
\label{eq:aLEFM_exact}
\end{equation}
Using Eq.~\eqref{eq:GLEFM}, this expression reduces to the brittle weak-spot length in the sharp-interface limit \(\delta\to0\),
\begin{equation}
a_{\mathrm{LEFM}}\to a_{c0}=\Lambda\frac{\tau_p-\tau_g}{\tau_g-\tau_r}.
\end{equation}

The relation between the present finite-softening weak-spot solution and the corresponding LEFM sharp-crack description can now be made explicit in the small-process-zone, or small-scale-yielding, limit \(\omega/\Lambda\ll1\). Since Eq.~\eqref{eq:omega_small_delta} gives
\[
\frac{\omega}{\Lambda}
\simeq
\frac{\tau_p}{\tau_p-\tau_g}\frac{\delta}{u_p},
\]
expanding Eq.~\eqref{eq:aLEFM_exact} gives
\begin{equation}
a_{\mathrm{LEFM}}
=
a_{c0}
\left[
1+\frac{\tau_p}{2(\tau_p-\tau_g)}\frac{\delta}{u_p}
\right]
+
O\!\left[\left(\frac{\delta}{u_p}\right)^2\right]
=
a_{c0}
\left[
1+\frac12\frac{\omega}{\Lambda}
\right]
+
O\!\left[\left(\frac{\omega}{\Lambda}\right)^2\right].
\label{eq:aLEFM_exp}
\end{equation}
By comparison, the finite-softening weak-spot solution gives, to first order in \(\delta/u_p\),
\begin{equation}
a_c
=
a_{c0}
\left[
1+\frac{\tau_p(\tau_p+\tau_r-2\tau_g)}{2(\tau_p-\tau_g)^2}\frac{\delta}{u_p}
\right]
+
O\!\left[\left(\frac{\delta}{u_p}\right)^2\right]
=
a_{c0}
\left[
1+
\frac{\tau_p+\tau_r-2\tau_g}{2(\tau_p-\tau_g)}
\frac{\omega}{\Lambda}
\right]
+
O\!\left[\left(\frac{\omega}{\Lambda}\right)^2\right],
\label{eq:ac_CZM_exp}
\end{equation}
and
\begin{equation}
b_c
=
a_{c0}
\left[
1+\frac{\tau_p(\tau_p-\tau_r)}{2(\tau_p-\tau_g)^2}\frac{\delta}{u_p}
\right]
+
O\!\left[\left(\frac{\delta}{u_p}\right)^2\right]
=
a_{c0}
\left[
1+
\frac{\tau_p-\tau_r}{2(\tau_p-\tau_g)}
\frac{\omega}{\Lambda}
\right]
+
O\!\left[\left(\frac{\omega}{\Lambda}\right)^2\right].
\label{eq:bc_CZM_exp}
\end{equation}
The three lengths coincide in the sharp-interface limit,
\[
a_c=b_c=a_{\mathrm{LEFM}}=a_{c0}\qquad\text{as}\qquad\delta\to0 .
\]
For a small but finite process zone, Eq.~\eqref{eq:aLEFM_exp} lies exactly midway between the rear and the front of the resolved process zone:
\begin{equation}
a_{\mathrm{LEFM}}=\frac12(a_c+b_c)+O\!\left[\left(\frac{\delta}{u_p}\right)^2\right]
=
\frac12(a_c+b_c)+O\!\left[\left(\frac{\omega}{\Lambda}\right)^2\right].
\label{eq:aLEFM_midpoint}
\end{equation}
Thus, the sharp-crack LEFM length should not be identified with either \(a_c\) or \(b_c\) individually. Rather, it represents an effective crack tip located inside the finite process zone resolved by the weak-spot formulation with finite interface softening. In this sense, \(a_c\) and \(b_c\) provide lower and upper mechanically resolved bounds for the equivalent LEFM crack length.

This result clarifies the connection with earlier fracture-energy interpretations (e.g. \cite{Gaume2014}). The brittle weak-spot length is recovered when the process zone vanishes. For finite softening, the weak-spot formulation with a finite-softening interface law provides more information than a sharp-crack LEFM description because it resolves both the fully residual crack length \(a_c\) and the total affected length \(b_c\). This distinction is important when comparing with numerical simulations or experiments, where the measured crack length may correspond either to the fully failed region or to the full damaged or process zone.

\section{Anticrack propagation}
\label{sec:anticrack_growth}

\subsection{Timoshenko simplified anticrack model}

Although the derivation above was restricted to slope-parallel shear failure, the same compliant-interface logic can be extended to collapse-driven anticrack propagation, where weak-layer compression and slab bending play the central role. Following the classical Timoshenko beam-on-Winkler-foundation framework \citep{Winkler1867,Timoshenko1921}, as adapted to weak-layer anticrack propagation by \citet{rosendahl2020modeling1,rosendahl2020modeling2,weissgraeber2023closed}, we use a Timoshenko beam resting on a compressive Winkler-type weak layer. The slope-normal deflection is denoted by \(w(x)\), and the cross-section rotation by \(\varphi(x)\). In this simplified anticrack analogue, the axial displacement \(u\) is not solved as part of the normal bending problem, so the \((w,\varphi)\) field is treated independently. For a homogeneous slab of thickness \(h\),
\[
\mathcal B=\frac{E'h^3}{12},\qquad
K_s=\kappa Gh,
\]
where \(\mathcal B\) is the bending stiffness and \(K_s\) is the transverse-shear rigidity per unit width. The Timoshenko resultants are
\[
M=\mathcal B\varphi',\qquad
V=K_s(w'+\varphi),
\]
so that slab shear deformation is retained explicitly.

In the intact part of the weak layer, the normal support is \(\sigma=k_n w\). The two characteristic quantities controlling the sharp-front Timoshenko anticrack are
\[
\Lambda_b=\left(\frac{4\mathcal B}{k_n}\right)^{1/4},
\qquad
\eta_B=\frac{\mathcal B}{K_s\Lambda_b^2},
\qquad
p_T=\sqrt{1+\eta_B}.
\]
The parameter \(\eta_B\) measures the importance of Timoshenko shear deformation relative to bending over the bending-foundation length. In the limit \(K_s\to\infty\), \(\eta_B\to0\) and \(p_T\to1\), recovering the classical bending-dominated result.

For a sharp collapsed zone of length \(a\), with residual normal support \(\sigma_r\), the Timoshenko matching gives the normal stress at the crack front as
\begin{equation}
\sigma_{\rm tip}
=
\sigma_g+
(\sigma_g-\sigma_r)
\left[
2p_T\frac{a}{\Lambda_b}
+
\left(\frac{a}{\Lambda_b}\right)^2
\right].
\label{eq:timo_sigma_tip_main}
\end{equation}
Thus, the brittle flat anticrack length follows from \(\sigma_{\rm tip}=\sigma_p\):
\begin{equation}
a_{c0}^{T}
=
\Lambda_b
\left[
\sqrt{
1+\eta_B+
\frac{\sigma_p-\sigma_g}{\sigma_g-\sigma_r}
}
-
\sqrt{1+\eta_B}
\right].
\label{eq:timo_ac0_main}
\end{equation}
For a void-like collapsed weak layer, \(\sigma_r=0\), this reduces to
\begin{equation}
a_{c0}^{T}
=
\Lambda_b
\left[
\sqrt{\frac{\sigma_p}{\sigma_g}+\eta_B}
-
\sqrt{1+\eta_B}
\right].
\label{eq:timo_ac0_void_main}
\end{equation}
Equations~\eqref{eq:timo_sigma_tip_main}--\eqref{eq:timo_ac0_void_main} are the compact Timoshenko sharp-front anticrack formula used here.

For finite compressive softening, the weak-layer support follows the same three-zone structure as in the shear problem: a fully collapsed zone, a linear softening zone, and an intact elastic zone. The normal softening stiffness is
\[
k_s=\frac{\sigma_p-\sigma_r}{\delta_n},
\]
and the associated bending-controlled softening length is
\[
\ell_b=
\left(\frac{\mathcal B}{k_s}\right)^{1/4}
=
\left(\frac{\mathcal B\delta_n}{\sigma_p-\sigma_r}\right)^{1/4}.
\]
In the Timoshenko formulation, the softening-zone roots also depend on the shear-deformation length
\[
\ell_s=\left(\frac{K_s}{k_s}\right)^{1/2}.
\]
The exact finite-softening problem therefore remains a linear constant-coefficient matching problem in each zone, but it contains both \(\ell_b\) and \(\ell_s\). The details are given in Appendix~\ref{app:flat_compressive_anticrack}. The important point for interpretation is that the bending-controlled \(\delta_n^{1/4}\) scale remains present, while Timoshenko shear deformation modifies the prefactors and the sharp-front length through \(\eta_B\).

\begin{figure}[t]
\centerline{\includegraphics[width=0.5\textwidth]{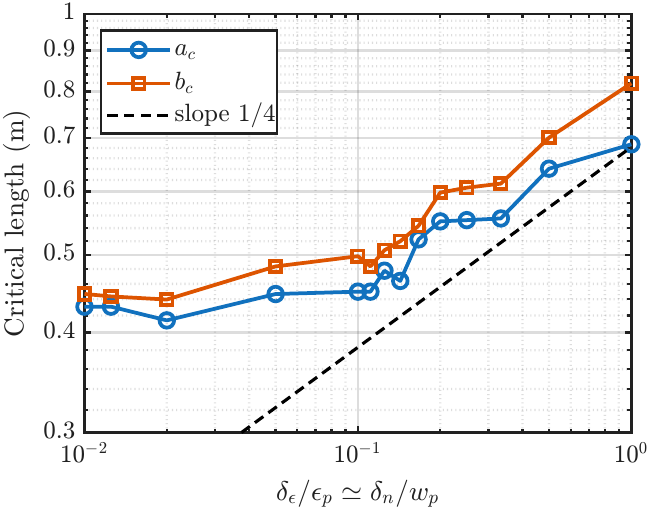}}
\caption{Effect of the normalized compressive softening distance on the characteristic anticrack lengths obtained from 3D MPM simulations based on the model of \protect\citet{Gaume2018}. The collapsed length \(a_c\) and the total affected length \(b_c\) are plotted as functions of \(\delta_\epsilon/\epsilon_p\simeq\delta_n/w_p\) in log--log scale ($\delta_\epsilon$ is the plastic strain and $\epsilon_p$ is the peak strain). Both lengths exhibit an approximately quarter-power dependence over the explored range, consistent with the bending-softening length \(\ell_b\sim(\mathcal{B}\delta_n/(\sigma_p-\sigma_r))^{1/4}\), which remains a controlling scale in the Timoshenko anticrack analogue. The dashed line indicates a reference slope of \(1/4\).}
\label{fig:anticrack_softening}
\end{figure}

The MPM simulations suggest that, over the explored range of \(\delta_n/w_p\), both the collapsed length \(a_c\) and the total affected length \(b_c\) exhibit an apparent quarter-power dependence on the compressive softening distance, while consistently approaching the brittle limit \(a_c=b_c=a_{c0}^{T}\) as \(\delta_n\to0\). Guided by the exact three-zone reduction and supported by the MPM scaling, a compact global approximation preserving the Timoshenko brittle limit is
\[
a_c(\delta_n)
\simeq
a_{c0}^{T}
\left(
1+C_a^{(b,T)}\frac{\delta_n}{w_p}
\right)^{1/4},
\]
where \(C_a^{(b,T)}\) is a dimensionless coefficient depending on the complete matching problem, including \(\eta_B\). This form should be understood as a compact representation of the finite-softening trend rather than as an independent fracture criterion.

The direct comparison between the MPM simulations and the present simplified interface-based anticrack model should nevertheless be interpreted with care. In an interface formulation, failure is necessarily localized across the full weak-layer thickness, whereas in a finite-thickness weak layer the deformation and damage may localize within only part of the layer. The effective interpretation of the softening distance \(\delta_n\) and of the elastic displacement \(w_p\) is therefore less direct, especially because the present MPM simulations are not regularized with respect to localization width. This may explain, for instance, why the simplified analytical model predicts a similar softening dependence for the total affected length \(b_c\) as the MPM simulations, while predicting a nearly constant or slightly decreasing collapsed length \(a_c\) with increasing \(\delta_n\). Such a trend for $a_c$ is not observed in the MPM results, likely because the measured damaged region is influenced by finite-thickness localization and by the absence of a regularized process-zone width.

\subsection{Mixed-mode anticrack model: simplified analytical and complete softening formulations}
\label{sec:mixed_mode_bending}

The flat anticrack analogue above describes the slope-normal response of the weak layer. On a slope, however, the weak layer may reach the onset of softening under combined compression and shear. In the reduced mixed-mode construction, the normal field is still obtained from the Timoshenko anticrack matching problem, while the shear stress at the leading edge of the affected zone is estimated from two contributions: amplification of the slope-parallel gravitational shear load and the tangential displacement generated by slab rotation. This reduced construction is only partially coupled: the Timoshenko normal problem provides \(w\) and \(\varphi\), whereas \(u\) is not solved as a separate field. Tangential loading is instead introduced through the shear-transfer estimate and the rotation-induced basal displacement, so that normal--shear coupling enters through \(\varphi\) and through the local mixed-mode peak condition.

For the sake of developing an elegant analytical formula, we use the sharp-crack limit. Thus the fracture-process zone is collapsed and there is only one front location,
\[
b_c=a_c .
\]
We therefore write all compact mixed-mode quantities in terms of \(a_c\). The distinction between \(a_c\) and \(b_c\) is retained only in the finite-softening appendix.

Consistently with the anticrack interpretation, the collapsed zone is treated as void-like in both normal and tangential directions,
\[
\sigma_r=0,
\qquad
\tau_r=0.
\]
The local stress state at the sharp crack front, \(x=a_c\), is denoted by
\[
\tau_{\rm tip}=\tau(a_c),
\qquad
\sigma_{\rm tip}=\sigma(a_c).
\]
Onset of local mixed-mode failure is described by the peak surface
\begin{equation}
\left(\frac{\tau_{\rm tip}}{\tau_p}\right)^2
+
\left(\frac{\sigma_{\rm tip}}{\sigma_p}\right)^2
=1.
\label{eq:mixed_yield_main_timo}
\end{equation}

The reduced mixed-mode correction is evaluated sequentially. First, we solve the slope-normal Timoshenko anticrack problem. In this step the unknown state is
\[
\bm y=[w,\varphi,M,V]^{\mathsf T},
\]
and the axial displacement \(u\) is not part of the boundary-value problem. Thus the normal bending problem is solved independently of the axial response. This gives both the normal stress amplification \(\sigma_{\rm tip}\) and the section rotation \(\varphi(a_c)\).

The rotation \(\varphi(a_c)\) is therefore not an additional empirical input. With \(\bm e_\varphi=(0,1,0,0)^{\mathsf T}\),
\[
\varphi(a_c)=\bm e_\varphi^{\mathsf T}\bm y_{I}(a_c)
=\bm e_\varphi^{\mathsf T}\bm y_{III}(a_c).
\]
In the sharp-front limit, this reduces to the compact expression
\[
|\varphi^{(0)}(a_c)|
=
2\,\frac{\sigma_g-\sigma_r}{k_n}
\frac{a_c}{\Lambda_b^2}
\left(1+p_T\frac{a_c}{\Lambda_b}\right),
\]
with the sign set by the slope and cutting convention.

Second, the tangential stress at the weak-layer front is reconstructed from two reduced contributions. The first is the slope-parallel shear-load amplification already present in the one-dimensional shear solution. The second is the basal tangential displacement induced by rotation of the slab cross-section. In Timoshenko kinematics this displacement is controlled by the section rotation \(\varphi\), not by imposing \(\varphi=w'\). We therefore write
\begin{equation}
u_{\rm bend}=-\frac{h}{2}\varphi(a_c),
\qquad
\tau_{\rm bend}^{T}=-k_t\frac{h}{2}\varphi(a_c),
\label{eq:timo_tau_bend_main}
\end{equation}
where \(k_t=\tau_p/u_p\) is the pre-peak tangential stiffness. The sign depends on the adopted shear convention; the calculations retain the signed value.

The shear stress entering the reduced mixed-mode criterion is then
\begin{equation}
\tau_{\rm tip}
=
\tau_g\left(1+\frac{a_c}{\Lambda}\right)
+
\tau_{\rm bend}^{T},
\qquad
\Lambda=\sqrt{\frac{E'h}{k_t}}.
\label{eq:timo_tau_tip_main}
\end{equation}
The first term represents shear-load amplification over the sharp-crack length, while the second term is the Timoshenko rotation-induced basal shear. Equation~\eqref{eq:timo_sigma_tip_main} gives the corresponding compact normal-stress reference
\[
\sigma_{\rm tip}
=
\sigma_g+
(\sigma_g-\sigma_r)
\left[
2p_T\frac{a_c}{\Lambda_b}
+
\left(\frac{a_c}{\Lambda_b}\right)^2
\right].
\]
The rotation-induced shear term can equivalently be written, in the sharp-front limit, as
\begin{equation}
\tau_{\rm bend}^{(0,T)}
=
\pm(\sigma_g-\sigma_r)\frac{k_t}{k_n}
\frac{h a_c}{\Lambda_b^2}
\left(1+p_T\frac{a_c}{\Lambda_b}\right),
\label{eq:timo_tau_bend_sharp_main}
\end{equation}
with the sign fixed by the slope and cutting convention. Thus the mixed-mode correction is now fully expressed in terms of Timoshenko quantities: \(\Lambda_b\), \(\eta_B\), \(p_T\), and the section rotation \(\varphi\).

The reduced mixed-mode model remains intentionally compact: the normal and tangential fields are combined at the leading edge through Eq.~\eqref{eq:mixed_yield_main_timo}, rather than by solving a full vectorial weak-layer boundary-value problem. Appendix~\ref{app:mixed_mode_anticrack} gives the finite-softening Timoshenko matching formulation. Appendix~\ref{app:fully_coupled_timoshenko} describes the fully coupled cohesive Timoshenko model, in which axial displacement, transverse deflection, section rotation, normal support, and tangential support are solved simultaneously. The sharp-crack limit of this formulation is used below as a reference solution. The same appendix also explains how an additional surface load and its possible eccentricity are included.

Thus, two levels of description are used. The reduced Timoshenko model provides a compact analytical correction in the sharp-front limit, while the fully coupled Timoshenko model solves the vectorial weak-layer problem directly. In the latter formulation, axial displacement, transverse deflection, section rotation, normal support, and tangential support are coupled in a single boundary-value problem. The fully coupled model is used below both in the sharp-crack limit and, with an associated mixed-mode cohesive law, in a finite-softening form. This finite-softening extension prescribes how the normal and tangential tractions degrade together after the mixed-mode peak surface is reached, and thereby resolves the process zone explicitly.

\subsection{Comparison with experimental data}
\label{sec:anticrack_data_comparison}

\begin{figure}[t]
\centerline{\includegraphics[width=0.6\textwidth]{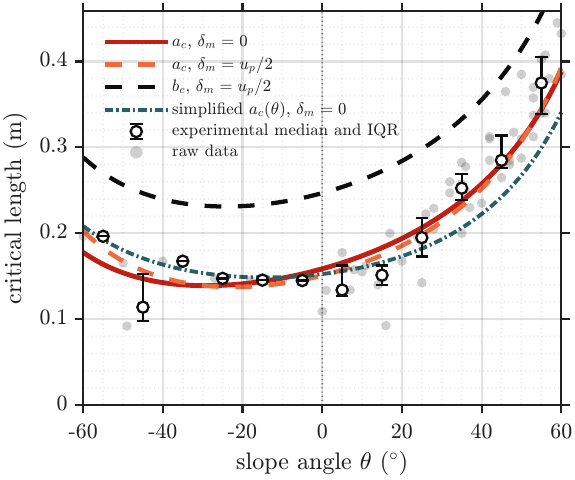}}
\caption{Mixed-mode anticrack model predictions and comparison with experiments. Critical length as a function of slope angle, \(\theta\). The fully coupled Timoshenko sharp-crack model is compared with the corresponding finite-softening solution and with the simplified analytical Timoshenko correction. Grey symbols show individual experimental measurements, while black symbols and error bars denote the experimental median and interquartile range (IQR) within \(10^\circ\) slope-angle bins.}
\label{fig:mixedmode}
\end{figure}

We now compare the mixed-mode anticrack models developed above with the critical cut-length measurements of \citet{adam2024fracture}. Three model predictions are shown in Fig.~\ref{fig:mixedmode}. The first is the simplified sharp-front analytical Timoshenko mixed-mode correction, shown by the dash-dotted green line. The second is the fully coupled Timoshenko sharp-crack model, corresponding to zero softening distance, shown by the solid red line. The third is the fully coupled finite-softening model with equal normalized normal and tangential softening distances,
\[
\frac{\delta_t}{u_p}
=
\frac{\delta_n}{w_p}
=
\frac12 .
\]
For this case, the red dashed line denotes the collapsed cut length \(a_c\), while the black dashed line denotes the total affected length \(b_c=a_c+\omega\), where \(\omega\) is the mixed-mode fracture-process-zone length.

The experimental observable is the imposed cut length. It should therefore be compared primarily with \(a_c\), which represents the fully collapsed or stress-free part of the weak layer. The quantity \(b_c\), by contrast, marks the leading edge of the finite-softening process zone. It is not directly equivalent to the visible cut length, but it provides useful information on the spatial extent of partial damage ahead of the cut.

The additional surface load applied in the experiments is accounted for separately from the slab thickness. Thus, it contributes to the gravitational loading but not to the slab bending stiffness. Following the load-decomposition strategy used in WEAC, we also account for the eccentricity of this surface load relative to the slab reference axis. The slab thickness is \(h=0.115~\mathrm{m}\), the density is \(\rho=350~\mathrm{kg\,m^{-3}}\), the Young's modulus is \(E=93~\mathrm{MPa}\), derived from density following \citet{adam2024fracture}, and the Poisson ratio is \(\nu=0.25\). We account for the additional surface load \(q_{\rm surf}=1.50~\mathrm{kPa}\), which is treated as fully eccentric as in \citet{adam2024fracture} (see Appendix~\ref{app:fully_coupled_timoshenko}).

The weak-layer thickness is \(h_w=0.010~\mathrm{m}\). The mixed-mode peak strengths are \(\sigma_p=5.5~\mathrm{kPa}\) in compression and \(\tau_p=4.0~\mathrm{kPa}\) in shear. These two strength values were calibrated to match the experimental critical-length data, but their magnitudes remain in good agreement with reported failure stresses for natural surface-hoar samples \citep{Melin2026}. We use effective weak-layer moduli over the thickness \(h_w\): a normal modulus \(E_{\rm wl}=0.20~\mathrm{MPa}\) and a shear modulus \(G_{\rm wl}=0.10~\mathrm{MPa}\). The resulting characteristic lengths are \(\Lambda_b=0.224~\mathrm{m}\) and \(\Lambda=1.068~\mathrm{m}\), with \(\eta_B=0.070\) and \(p_T=\sqrt{1+\eta_B}=1.035\).

Overall, all three models reproduce the main experimental trend: the critical cut length is smallest close to horizontal conditions, increases rapidly with slope angle on the positive branch, and also increases more moderately as the absolute value of the slope angle grows on the negative branch. The resulting dependence is therefore asymmetric and broadly U-shaped, with a much steeper rise for positive than for negative slope angles. The simplified analytical model captures this overall shape, but tends to underpredict the rapid increase in critical length at large positive slope angles. The \(R^2\) values reported here are computed by interpolating each model curve at the experimental slope angles and comparing the predictions with the individual measured cut lengths. The quantitative agreement of the simplified analytical model is weaker, with \(R^2\simeq0.78\). The fully coupled sharp-crack limit improves the match substantially, with \(R^2\simeq0.90\), indicating that coupled normal--tangential weak-layer response, slab rotation, and load eccentricity are important for reproducing the measured slope-angle dependence.

Including finite softening with \(\delta_t/u_p=\delta_n/w_p=1/2\) only weakly modifies the predicted cut length \(a_c\) relative to the sharp-crack fully coupled solution, but it slightly improves the agreement with the experimental data, yielding \(R^2\simeq0.92\). Thus, for the present parameters, finite softening does not strongly modify the measured critical cut length. However, the same calculation predicts a non-negligible fracture process zone ahead of the cut, of order \(10~\mathrm{cm}\), so that the total affected length \(b_c\) is much larger than \(a_c\). This shows that the experimentally measured cut length can remain close to the brittle prediction even when a partially damaged zone exists ahead of the cut.

The comparison therefore suggests that realistic elastic properties, together with a mixed-mode weak-layer failure envelope constrained by laboratory measurements, are sufficient to reproduce both the magnitude and the slope-angle dependence of the critical cut length. At the same time, the difference between \(a_c\) and \(b_c\) highlights the importance of distinguishing the imposed cut length from the full process-zone length when finite softening is considered.

\subsubsection{Relation to WEAC and equivalent toughness}
\label{sec:weac_connection}

The fully coupled Timoshenko formulation used here is closely related to WEAC-type models. In both approaches, the slab is described as a shear-deformable beam resting on a weak-layer foundation with normal and tangential stiffnesses. The mechanical state is governed by the same six-field vector,
\[
\bm z=[u,u',w,w',\varphi,\varphi']^{\mathsf T},
\]
where \(u\) is the axial displacement, \(w\) the transverse deflection, and \(\varphi\) the section rotation. Both formulations use the laminate stiffness resultants \(A_{11}\), \(B_{11}\), \(D_{11}\), and \(kA_{55}\), and the same type of free and supported beam segments connected by continuity of displacement and internal resultants. The treatment of additional surface loading and its eccentricity also follows the same mechanical logic as in WEAC.

The main difference lies in the propagation criterion. WEAC evaluates anticrack propagation in a sharp-crack fracture-mechanics framework, by computing mode-I and mode-II energy-release rates and comparing them with prescribed fracture toughnesses. In the notation used here, the corresponding local energy-release-rate measures are
\[
G_I=\frac12 k_n w_{\rm tip}^2
=
\frac{\sigma_{\rm tip}^2}{2k_n},
\qquad
G_{II}
=
\frac{\tau_{\rm tip}^2}{2k_t}.
\]
Thus, WEAC requires \(G_{Ic}\) and \(G_{IIc}\) as material inputs. In contrast, the present model prescribes the local weak-layer stress--displacement law. The fracture process zone is therefore resolved explicitly, and the equivalent toughness follows from the area under the constitutive law rather than being imposed independently.

For the void-like collapsed weak layer considered here, \(\sigma_r=\tau_r=0\). The brittle, or zero-softening, equivalent pure-mode toughnesses are therefore
\[
G_{Ic}^{(0)}
=
\frac12\sigma_p w_p
=
\frac{\sigma_p^2}{2k_n},
\qquad
G_{IIc}^{(0)}
=
\frac12\tau_p u_p
=
\frac{\tau_p^2}{2k_t}.
\]
With the parameters used in Fig.~\ref{fig:mixedmode}, we get
\[
G_{Ic}^{(0)}=0.76~\mathrm{J\,m^{-2}},
\qquad
G_{IIc}^{(0)}=0.80~\mathrm{J\,m^{-2}}.
\]
These values are in excellent agreement with the mixed-mode anticrack toughness values reported by \citet{adam2024fracture}. Thus, the weak-layer strength and stiffness values used in the present model imply equivalent fracture toughnesses, without prescribing toughness as an independent input.

When finite softening is included, the equivalent toughness increases by the post-peak work. For the linear softening law and zero residual strength,
\[
G_{Ic}
=
\frac12\sigma_p w_p
+
\frac12\sigma_p\delta_n,
\qquad
G_{IIc}
=
\frac12\tau_p u_p
+
\frac12\tau_p\delta_t .
\]
For the finite-softening calculation shown in Fig.~\ref{fig:mixedmode}, we use equal normalized normal and tangential softening distances,
\[
\frac{\delta_t}{u_p}
=
\frac{\delta_n}{w_p}
=
\frac12,
\]
which gives
\[
G_{Ic}=1.13~\mathrm{J\,m^{-2}},
\qquad
G_{IIc}=1.20~\mathrm{J\,m^{-2}}.
\]

More generally, if residual stresses are retained, the corresponding effective sharp-crack toughnesses are obtained by integrating the cohesive work above the residual branch,
\[
G_{Ic}^{\rm eff}
=
\int_0^{w_r}\left[\sigma(w)-\sigma_r\right]\,dw,
\qquad
G_{IIc}^{\rm eff}
=
\int_0^{u_r}\left[\tau(u)-\tau_r\right]\,du .
\]
For a piecewise linear law, this gives
\[
G_{Ic}^{\rm eff}
=
\left(\frac12\sigma_p-\sigma_r\right)w_p
+
\frac12(\sigma_p-\sigma_r)\delta_n,
\]
and
\[
G_{IIc}^{\rm eff}
=
\left(\frac12\tau_p-\tau_r\right)u_p
+
\frac12(\tau_p-\tau_r)\delta_t .
\]
The present formulation can therefore be interpreted as a cohesive-zone regularization of a WEAC-like sharp-crack model. In the limit \(\delta_t,\delta_n\to0\), the process zone collapses and the model approaches a sharp-front description. In that limit, the present model becomes equivalent to a WEAC-type formulation provided the critical energy-release rates are identified with the elastic energy stored in the weak-layer springs at peak stress, \(G_{I0}=\sigma_p^2/(2k_n)\) and \(G_{II0}=\tau_p^2/(2k_t)\). For the quadratic mixed-mode peak-stress envelope used here, this leads directly to a linear interaction in energy-release-rate space,
\[
\frac{G_I}{G_{I0}}+\frac{G_{II}}{G_{II0}}=1,
\]
that is, to a linear decrease of \(G_{II}\) with increasing \(G_I\). This interaction is broadly consistent with the scatter of the mixed-mode toughness data of \citet{adam2024fracture}, although their WEAC analysis introduced additional empirical exponents to tune the shape of the toughness envelope. In the present formulation, this shape is not imposed independently at the toughness level, but follows from the assumed weak-layer failure envelope and elastic constants. Additional flexibility could therefore be introduced, if needed, by enriching the mixed-mode failure envelope itself rather than by prescribing separate toughness-interaction exponents.

For non-zero softening distances, however, the process zone is retained explicitly, which allows one to distinguish the collapsed length \(a_c\) from the total affected length \(b_c\). The central difference with a toughness-based sharp-crack formulation is therefore that fracture toughness is not required as an independent model input. Instead, the equivalent toughness is derived a posteriori from weak-layer strength, stiffness, residual resistance, and softening distance. This is potentially advantageous for snowpack and avalanche-forecasting applications, where strength and elastic properties are more readily parameterized, often through density-based relationships, than fracture toughness.

\section{Conclusion}

We derived an analytical model of snow slab avalanche release in which crack propagation is governed by shear failure in a weak layer beneath an elastic snow slab, explicitly accounting for both weak-layer pre-peak compliance and finite post-peak softening. The model recovers the classical brittle weak-spot critical length in the limit of vanishing softening distance and provides closed-form expressions for the fully softened crack length \(a_c\) and for the total affected length \(b_c\), which also includes the fracture process zone. Our results highlight the need for caution when interpreting critical lengths from numerical simulations or experiments: the fully softened or residual crack length should be distinguished from the full affected length, which may include partially damaged material ahead of the visible crack.

The exact shear solution admits a compact square-root expression for \(a_c\), whose small-softening limit is linear in \(\delta/u_p\), while the fracture-process-zone length \(\omega=b_c-a_c\) also grows linearly with \(\delta/u_p\). The formulation provides a direct connection between weak-spot models and sharp-crack fracture-energy descriptions. In the present approach, however, fracture energy is not imposed as an external propagation criterion; it emerges from the local stress--displacement law of the weak layer. The corresponding LEFM crack length can then be interpreted as an effective sharp-crack representation of a finite process zone resolved explicitly by the softening model.

We also extended the same compliant-softening framework to collapse-driven anticrack propagation. The simplified Timoshenko anticrack analogue shows that slab bending introduces a bending-controlled length scale and an approximately \(\delta^{1/4}\)-type dependence on the compressive softening distance, distinct from the shear-failure scaling. The mixed-mode Timoshenko formulation further shows that weak-layer compression, slope-parallel shear, slab rotation, and load eccentricity can be combined to reproduce the observed dependence of critical cut length on slope angle. The fully coupled version of the model is close in spirit to WEAC-type beam-on-foundation formulations, since it solves the coupled Timoshenko slab response on a weak-layer foundation. The main difference is conceptual: no fracture toughness needs to be prescribed as an independent model input. Instead, an effective toughness or fracture energy can be derived a posteriori from the weak-layer constitutive law.

This feature may be useful for operational snowpack modeling. Strength and elastic properties of snow layers are more readily parameterized in snowpack models, often through density-based relationships, than fracture toughness. A formulation based directly on weak-layer strength, stiffness, residual resistance, and softening distance may therefore offer a natural route for integration into avalanche forecasting models. Overall, the present work revisits classical shear-failure avalanche-release theory in a more complete mechanical framework, connects it to anticrack and mixed-mode propagation, and provides compact benchmarks for numerical models, laboratory experiments, and future operational parameterizations.

\paragraph{Acknowledgements}

The authors gratefully acknowledge the researchers involved in the discussion of the paper entitled ``Parameterization of the snow fracture energy to model the onset of crack propagation in snowpack models'' in \textit{The Cryosphere Discussions}, whose comments and exchanges motivated the present work.
The authors acknowledge the use of artificial intelligence as a limited exploratory tool during the algebraic analysis. Motivated by the DAMPM results, which suggested that the effect of finite softening on the fully residual crack length could be represented by a square-root correction, AI was consulted to assess whether such a compact reformulation was algebraically plausible before carrying out the derivation from the exact matching conditions. This use was restricted to exploring a concise representation of an already closed analytical solution; the square-root reformulation is not required for the validity of the model, but provides a useful compact form for interpretation and comparison.

\paragraph{Authors' contributions}
J.G. conceived the study and developed the analytical framework. F.M. contributed to the numerical simulations and model comparison. I.R. contributed experimental data and interpretation of weak-layer shear tests. P.H. contributed to the fracture-energy interpretation and to the connection with the equivalent LEFM formulation. All authors discussed the results and contributed to the manuscript.

\paragraph{Competing interests}
The authors declare that they have no competing interests.

\appendix

\section{Shear model: Derivation of the exact matching conditions}
\label{app:matching}

This appendix gives the details leading to Eqs.~\eqref{eq:alpha_eq}--\eqref{eq:bc_closed}. We use the notation \(a_c\) for the fully residual half-length, \(b_c\) for the total affected half-length, and
\[
\omega=b_c-a_c,
\qquad
\alpha=\frac{\omega}{\ell}.
\]

\subsection{Intact zone}

In the intact zone, \(x\ge b_c\), the interface is elastic:
\[
\tau(u)=\frac{\tau_p}{u_p}u .
\]
Equation~\eqref{eq:gov} becomes
\begin{equation}
E'h\,u''-\frac{\tau_p}{u_p}u=-\tau_g .
\end{equation}
Using
\[
\Lambda=\sqrt{\frac{E'h\,u_p}{\tau_p}},
\]
the bounded solution as \(x\to+\infty\) is
\begin{equation}
u_{III}(x)
=
\frac{\tau_g}{\tau_p}u_p
+
u_p\frac{\tau_p-\tau_g}{\tau_p}
e^{-(x-b_c)/\Lambda}.
\label{eq:app_uIII}
\end{equation}
Its derivative at \(x=b_c\) is
\begin{equation}
u'_{III}(b_c)
=
-\frac{u_p}{\Lambda}\frac{\tau_p-\tau_g}{\tau_p}
=
-\frac{\Lambda}{E'h}(\tau_p-\tau_g).
\label{eq:app_uIIIprime}
\end{equation}

\subsection{Residual zone}

In the residual zone, \(0\le x\le a_c\), the interface carries the constant residual stress \(\tau_r\). Equation~\eqref{eq:gov} gives
\begin{equation}
E'h\,u''=\tau_r-\tau_g .
\end{equation}
With the symmetry condition \(u'(0)=0\), we obtain
\begin{equation}
u'_I(x)=\frac{\tau_r-\tau_g}{E'h}x .
\end{equation}
Therefore, at \(x=a_c\),
\begin{equation}
u'_I(a_c)=\frac{\tau_r-\tau_g}{E'h}a_c .
\label{eq:app_uIprime}
\end{equation}
The displacement condition at the residual--softening boundary is
\begin{equation}
u_I(a_c)=u_r=u_p+\delta .
\end{equation}

\subsection{Softening zone}

In the softening zone, \(a_c\le x\le b_c\), the interface follows
\[
\tau(u)=\tau_p-\frac{\tau_p-\tau_r}{\delta}(u-u_p).
\]
The governing equation becomes
\begin{equation}
E'h\,u''
+
\frac{\tau_p-\tau_r}{\delta}u
=
\tau_p+\frac{\tau_p-\tau_r}{\delta}u_p-\tau_g .
\end{equation}
Using
\[
\ell=\sqrt{\frac{E'h\,\delta}{\tau_p-\tau_r}},
\]
the solution can be written as
\begin{equation}
u_{II}(x)
=
u_c
+
A\cos\left(\frac{x-a_c}{\ell}\right)
+
B\sin\left(\frac{x-a_c}{\ell}\right),
\end{equation}
with
\begin{equation}
u_c
=
u_p+\delta\frac{\tau_p-\tau_g}{\tau_p-\tau_r}.
\end{equation}

Continuity of displacement at \(x=a_c\) gives
\begin{equation}
A
=
\delta\frac{\tau_g-\tau_r}{\tau_p-\tau_r}.
\end{equation}
Continuity of the derivative at \(x=a_c\), together with Eq.~\eqref{eq:app_uIprime}, gives
\begin{equation}
B
=
-\frac{\delta(\tau_g-\tau_r)}{\tau_p-\tau_r}
\frac{a_c}{\ell}.
\end{equation}
Thus,
\begin{equation}
u_{II}(x)
=
u_c
+
\delta\frac{\tau_g-\tau_r}{\tau_p-\tau_r}
\cos\left(\frac{x-a_c}{\ell}\right)
-
\delta\frac{\tau_g-\tau_r}{\tau_p-\tau_r}
\frac{a_c}{\ell}
\sin\left(\frac{x-a_c}{\ell}\right).
\label{eq:app_uII}
\end{equation}

\subsection{\boldmath Matching at \(x=b_c\)}

At \(x=b_c\), the displacement reaches the peak value \(u_p\). Using Eq.~\eqref{eq:app_uII} and \(\alpha=(b_c-a_c)/\ell\), the displacement condition gives
\begin{equation}
(\tau_p-\tau_g)
+
(\tau_g-\tau_r)\cos\alpha
-
(\tau_g-\tau_r)\frac{a_c}{\ell}\sin\alpha
=
0 .
\label{eq:app_match1}
\end{equation}
The derivative condition \(u'_{II}(b_c)=u'_{III}(b_c)\), using Eq.~\eqref{eq:app_uIIIprime}, gives
\begin{equation}
(\tau_g-\tau_r)
\left(
\ell\sin\alpha+a_c\cos\alpha
\right)
=
(\tau_p-\tau_g)\Lambda .
\label{eq:app_match2}
\end{equation}

Solving Eq.~\eqref{eq:app_match1} for \(a_c/\ell\) yields
\begin{equation}
\frac{a_c}{\ell}
=
\frac{
\dfrac{\tau_p-\tau_g}{\tau_g-\tau_r}
+
\cos\alpha
}{
\sin\alpha
}.
\label{eq:app_ac_over_ell}
\end{equation}
Substitution into Eq.~\eqref{eq:app_match2} gives
\begin{equation}
(\tau_g-\tau_r)
+
(\tau_p-\tau_g)\cos\alpha
=
(\tau_p-\tau_g)\frac{\Lambda}{\ell}\sin\alpha ,
\end{equation}
which is Eq.~\eqref{eq:alpha_eq}. Equations~\eqref{eq:ac_closed} and~\eqref{eq:bc_closed} then follow directly from Eq.~\eqref{eq:app_ac_over_ell} and \(b_c=a_c+\ell\alpha\).

\section{Shear model: Exact square-root reformulation and small-softening expansion}
\label{app:small_delta}

This appendix shows how the exact expression for the fully residual crack length
can be rewritten in square-root form, and then derives the linear small-softening
approximations for the fully residual length, the process-zone length, and the
total affected length.

We introduce
\begin{equation}
T=\tau_p-\tau_g,
\qquad
R=\tau_g-\tau_r,
\qquad
s=\frac{\ell}{\Lambda}.
\end{equation}
Equation~\eqref{eq:alpha_eq} becomes
\begin{equation}
R+T\cos\alpha=\frac{T}{s}\sin\alpha .
\label{eq:app_alpha_s}
\end{equation}
Dividing by \(R\) gives
\begin{equation}
1+m\cos\alpha=\frac{m}{s}\sin\alpha,
\qquad
m=\frac{T}{R}.
\label{eq:app_alpha_m}
\end{equation}
From Eq.~\eqref{eq:ac_closed},
\begin{equation}
\frac{a_c}{\ell}
=
\frac{m+\cos\alpha}{\sin\alpha}.
\label{eq:app_ac_over_l}
\end{equation}

We now eliminate \(\alpha\). Let
\[
q=\frac{a_c}{\ell}.
\]
Equation~\eqref{eq:app_ac_over_l} gives
\[
q\sin\alpha=m+\cos\alpha .
\]
Equation~\eqref{eq:app_alpha_m} gives
\[
\frac{m}{s}\sin\alpha=1+m\cos\alpha .
\]
Combining these two relations with \(\sin^2\alpha+\cos^2\alpha=1\) yields
\begin{equation}
q^2
=
m^2\left(1+\frac{1}{s^2}\right)-1 .
\label{eq:app_ac_exact_intermediate}
\end{equation}
Since \(q=a_c/\ell\), this gives
\begin{equation}
a_c^2
=
\ell^2
\left[
m^2\left(1+\frac{1}{s^2}\right)-1
\right]
=
\Lambda^2 m^2+\ell^2(m^2-1).
\label{eq:app_ac_exact_lam_ell}
\end{equation}
The brittle length is
\[
a_{c0}
=
\Lambda m
=
\Lambda\frac{\tau_p-\tau_g}{\tau_g-\tau_r}.
\]
Therefore,
\begin{equation}
a_c^2
=
a_{c0}^2
+
\ell^2
\left[
\left(\frac{\tau_p-\tau_g}{\tau_g-\tau_r}\right)^2-1
\right].
\label{eq:app_ac_exact_before_delta}
\end{equation}
Using
\[
\frac{\ell^2}{\Lambda^2}
=
\frac{\tau_p}{\tau_p-\tau_r}\frac{\delta}{u_p},
\]
and writing the result in terms of \(a_{c0}\), we obtain
\begin{equation}
a_c
=
a_{c0}
\sqrt{
1+
C_a\frac{\delta}{u_p}
},
\label{eq:app_ac_exact_sqrt}
\end{equation}
where
\begin{equation}
C_a
=
\frac{
\tau_p(\tau_p-2\tau_g+\tau_r)
}{
(\tau_p-\tau_g)^2
}.
\end{equation}
This proves Eq.~\eqref{eq:ac_sqrt_exact}. The square-root dependence of \(a_c\)
on \(\delta/u_p\) is therefore an exact reformulation of the closed-form solution,
not a small-softening approximation.

For small \(\delta/u_p\), Eq.~\eqref{eq:app_ac_exact_sqrt} can be linearized using
\(\sqrt{1+x}=1+x/2+O(x^2)\). This gives
\begin{equation}
a_c
=
a_{c0}
\left(
1+\frac{C_a}{2}\frac{\delta}{u_p}
\right)
+
O\!\left[
\left(\frac{\delta}{u_p}\right)^2
\right].
\label{eq:app_ac_linear}
\end{equation}

The process-zone length is
\[
\omega=\ell\alpha .
\]
For \(s\ll1\), we seek
\[
\alpha=m_1s+O(s^3).
\]
Substituting this expansion into Eq.~\eqref{eq:app_alpha_s} gives
\[
m_1=\frac{T+R}{T}
=
\frac{\tau_p-\tau_r}{\tau_p-\tau_g}.
\]
Therefore,
\[
\omega=\ell\alpha
\simeq
\Lambda s\,m_1s
=
\Lambda m_1s^2.
\]
Using
\[
s^2
=
\frac{\ell^2}{\Lambda^2}
=
\frac{\tau_p}{\tau_p-\tau_r}\frac{\delta}{u_p},
\]
we obtain
\begin{equation}
\omega
\simeq
\Lambda
\frac{\tau_p}{\tau_p-\tau_g}
\frac{\delta}{u_p}.
\label{eq:app_omega_small}
\end{equation}

Finally, the total affected length is
\[
b_c=a_c+\omega .
\]
Combining Eqs.~\eqref{eq:app_ac_linear} and \eqref{eq:app_omega_small} gives
\[
b_c
=
a_{c0}
+
\left[
\frac{a_{c0}C_a}{2}
+
\Lambda\frac{\tau_p}{\tau_p-\tau_g}
\right]
\frac{\delta}{u_p}
+
O\!\left[
\left(\frac{\delta}{u_p}\right)^2
\right].
\]
Using
\[
a_{c0}
=
\Lambda\frac{\tau_p-\tau_g}{\tau_g-\tau_r}
\]
and
\[
C_a
=
\frac{
\tau_p(\tau_p-2\tau_g+\tau_r)
}{
(\tau_p-\tau_g)^2
},
\]
this becomes
\begin{equation}
b_c
=
a_{c0}
+
\frac{
\Lambda\tau_p(\tau_p-\tau_r)
}{
2(\tau_p-\tau_g)(\tau_g-\tau_r)
}
\frac{\delta}{u_p}
+
O\!\left[
\left(\frac{\delta}{u_p}\right)^2
\right].
\label{eq:app_bc_first_order}
\end{equation}
Equivalently,
\begin{equation}
b_c
=
a_{c0}
\left(
1+\frac{C_b}{2}\frac{\delta}{u_p}
\right)
+
O\!\left[
\left(\frac{\delta}{u_p}\right)^2
\right],
\label{eq:app_bc_linear}
\end{equation}
with
\begin{equation}
C_b
=
\frac{
\tau_p(\tau_p-\tau_r)
}{
(\tau_p-\tau_g)^2
}.
\end{equation}
Thus, \(a_c\) admits an exact square-root expression, whereas \(a_c\), \(\omega\),
and \(b_c\) admit simple linear forms in the small-softening limit.

\section{Timoshenko anticrack analogue solution on flat terrain}
\label{app:flat_compressive_anticrack}

This appendix gives the Timoshenko version of the flat compressive anticrack analogue used in the main text. The slab is described by the slope-normal deflection \(w(x)\) and by the independent section rotation \(\varphi(x)\). For a homogeneous slab of unit width,
\[
\mathcal B=\frac{E'h^3}{12},
\qquad
K_s=\kappa Gh,
\]
where \(\mathcal B\) is the bending stiffness and \(K_s\) is the transverse-shear rigidity. The Timoshenko resultants are
\[
M=\mathcal B\varphi',
\qquad
V=K_s(w'+\varphi).
\]
The weak-layer compressive support law is
\[
\sigma(w)=
\begin{cases}
k_n w, & 0\leq w\leq w_p,\\[1ex]
\sigma_p-\dfrac{\sigma_p-\sigma_r}{\delta_n}(w-w_p),
& w_p\leq w\leq w_r,\\[2ex]
\sigma_r, & w\geq w_r,
\end{cases}
\]
with
\[
\sigma_p=k_n w_p,
\qquad
w_r=w_p+\delta_n,
\qquad
\Delta\sigma=\sigma_p-\sigma_r,
\qquad
k_s=\frac{\Delta\sigma}{\delta_n}.
\]

\subsection{First-order Timoshenko form}

Using the state vector
\[
\bm y=
\begin{bmatrix}
w & \varphi & M & V
\end{bmatrix}^{\mathsf T},
\]
the equilibrium equations can be written as
\[
w'=\frac{V}{K_s}-\varphi,
\qquad
\varphi'=\frac{M}{\mathcal B},
\qquad
M'=V,
\qquad
V'=\sigma(w)-\sigma_g .
\]
In any zone where the support law can be written as
\[
\sigma(w)=\sigma_0+k w,
\]
the state equation is linear with constant coefficients,
\begin{equation}
\bm y'=\bm A(k)\bm y+\bm d(\sigma_0),
\label{eq:timo_y_state}
\end{equation}
where
\[
\bm A(k)=
\begin{bmatrix}
0&-1&0&1/K_s\\
0&0&1/\mathcal B&0\\
0&0&0&1\\
k&0&0&0
\end{bmatrix},
\qquad
\bm d(\sigma_0)=
\begin{bmatrix}
0\\0\\0\\ \sigma_0-\sigma_g
\end{bmatrix}.
\]
The three zones correspond to
\[
\text{collapsed: } k=0,\quad \sigma_0=\sigma_r,
\]
\[
\text{softening: } k=-k_s,
\quad \sigma_0=\sigma_p+k_s w_p,
\]
\[
\text{intact: } k=k_n,
\quad \sigma_0=0.
\]
Thus, in each zone,
\begin{equation}
\bm y_j(x)=\exp[\bm A_j(x-x_j)]\bm c_j+\bm y_{p,j},
\label{eq:timo_zone_solution}
\end{equation}
where \(\bm y_{p,j}\) is a constant particular solution. In the collapsed zone, Eq.~\eqref{eq:timo_zone_solution} reduces to the corresponding polynomial Timoshenko beam solution. In the softening zone, the eigenvalues depend on both
\[
\ell_b=\left(\frac{\mathcal B}{k_s}\right)^{1/4},
\qquad
\ell_s=\left(\frac{K_s}{k_s}\right)^{1/2}.
\]
Therefore the finite-softening Timoshenko solution contains the bending-controlled length \(\ell_b\), responsible for the quarter-power dependence, and the shear-deformation length \(\ell_s\), which modifies the matching coefficients.

\subsection{Three-zone matching}

The zones are
\[
0\le x\le a_c,
\qquad
a_c\le x\le b_c,
\qquad
x\ge b_c,
\]
representing the collapsed, softening, and intact regions, respectively. At the symmetry/free end of the one-sided construction,
\[
M_I(0)=0,
\qquad
V_I(0)=0.
\]
At \(x=a_c\), the weak layer reaches the end of the softening branch,
\[
w(a_c)=w_r=w_p+\delta_n,
\]
and at \(x=b_c\), it reaches the onset of softening,
\[
w(b_c)=w_p.
\]
Across both interfaces the Timoshenko state is continuous:
\[
\bm y_I(a_c)=\bm y_{II}(a_c),
\qquad
\bm y_{II}(b_c)=\bm y_{III}(b_c).
\]
In the intact region the bounded modes decay toward the far-field state
\[
w_\infty=\frac{\sigma_g}{k_n},
\qquad
\varphi_\infty=0,
\qquad
M_\infty=0,
\qquad
V_\infty=0.
\]
The finite-softening anticrack problem is therefore reduced to a finite nonlinear matching system for \(a_c\), \(b_c\), and the remaining integration constants. This is the direct Timoshenko counterpart of the three-zone shear solution: the fields are analytical within each zone, while the unknown front locations are obtained from the matching conditions.

\subsection{Brittle sharp-front limit}

In the brittle limit \(\delta_n\to0\), the softening zone collapses and \(b_c\to a_c\). The intact Timoshenko foundation problem is controlled by
\[
\Lambda_b=\left(\frac{4\mathcal B}{k_n}\right)^{1/4},
\qquad
\eta_B=\frac{\mathcal B}{K_s\Lambda_b^2},
\qquad
p_T=\sqrt{1+\eta_B}.
\]
For a collapsed zone of length \(a\), direct matching between the unsupported and intact Timoshenko solutions gives
\begin{equation}
\sigma_{\rm tip}
=
\sigma_g+
(\sigma_g-\sigma_r)
\left[
2p_T\frac{a}{\Lambda_b}
+
\left(\frac{a}{\Lambda_b}\right)^2
\right].
\label{eq:timo_sigma_tip_app}
\end{equation}
Setting \(\sigma_{\rm tip}=\sigma_p\) yields
\begin{equation}
a_{c0}^{T}
=
\Lambda_b
\left[
\sqrt{
1+\eta_B+
\frac{\sigma_p-\sigma_g}{\sigma_g-\sigma_r}
}
-
\sqrt{1+\eta_B}
\right].
\label{eq:timo_ac0_app}
\end{equation}
For a void-like collapsed zone, \(\sigma_r=0\), this becomes
\begin{equation}
a_{c0}^{T}
=
\Lambda_b
\left[
\sqrt{\frac{\sigma_p}{\sigma_g}+\eta_B}
-
\sqrt{1+\eta_B}
\right].
\label{eq:timo_ac0_void_app}
\end{equation}
The role of \(\eta_B\) is therefore transparent in the sharp-front Timoshenko analogue: shear deformation decreases the stress amplification for a given \(a/\Lambda_b\), and therefore modifies the critical length through \(p_T=\sqrt{1+\eta_B}\).

\section{Finite-softening mixed-mode Timoshenko anticrack criterion}
\label{app:mixed_mode_anticrack}

This appendix gives the reduced mixed-mode Timoshenko correction used in the main text. The purpose is to account for local weak-layer softening under combined compression and shear while retaining a compact analytical structure. The model is still reduced: the normal anticrack field and the tangential shear-transfer estimate are combined through a local peak surface. A fully coupled sharp-crack formulation is described in Appendix~\ref{app:fully_coupled_timoshenko}.

We use the following convention. In the finite-softening problem, \(a_c\) denotes the end of the fully collapsed zone, whereas \(b_c\) denotes the leading edge of the process zone where the weak layer first reaches the mixed-mode peak surface. Therefore stresses entering the peak condition are evaluated at \(x=b_c\). In the sharp-crack limit the process-zone length vanishes, so \(b_c=a_c\). The compact mixed-mode correction in the main text uses this sharp-crack convention and is therefore written only with \(a_c\).

\subsection{Mixed-mode peak surface}

At the leading edge of the affected zone, \(x=b_c\), the local stress state is
\[
\sigma_{\rm tip}=\sigma(b_c),
\qquad
\tau_{\rm tip}=\tau(b_c).
\]
The onset of softening is assumed to occur on the elliptical peak surface
\begin{equation}
\left(\frac{\tau_{\rm tip}}{\tau_p}\right)^2
+
\left(\frac{\sigma_{\rm tip}}{\sigma_p}\right)^2
=1.
\label{eq:timo_mixed_peak_app}
\end{equation}
For the void-like anticrack construction,
\[
\sigma_r=0,
\qquad
\tau_r=0.
\]
Thus the collapsed zone carries neither normal support nor residual tangential resistance.

\subsection{Normal finite-softening field with unknown leading-edge stress}

In mixed mode, the normal stress at the onset of softening is not known a priori. It is denoted by \(\sigma_{\rm tip}\), with corresponding displacement
\[
w_{\rm tip}=\frac{\sigma_{\rm tip}}{k_n}.
\]
The normal softening branch from \(\sigma_{\rm tip}\) to zero over the distance \(\delta_n\) has tangent stiffness
\[
k_s^{\rm tip}=\frac{\sigma_{\rm tip}}{\delta_n},
\]
and therefore
\[
\ell_b^{\rm tip}=
\left(\frac{\mathcal B\delta_n}{\sigma_{\rm tip}}\right)^{1/4},
\qquad
\ell_s^{\rm tip}=
\left(\frac{K_s\delta_n}{\sigma_{\rm tip}}\right)^{1/2}.
\]
The Timoshenko state equation \eqref{eq:timo_y_state} is used in the collapsed, softening, and intact zones with \(w(b_c)=w_{\rm tip}\) and \(w(a_c)=w_{\rm tip}+\delta_n\). Solving the matching conditions gives the front rotation \(\varphi(b_c)\) and the two lengths \(a_c\) and \(b_c\) for any trial value of \(\sigma_{\rm tip}\). Explicitly, if \(\bm e_\varphi=(0,1,0,0)^{\mathsf T}\), then
\begin{equation}
\varphi(b_c)
=
\bm e_\varphi^{\mathsf T}\bm y_{II}(b_c)
=
\bm e_\varphi^{\mathsf T}
\left[
\exp(\bm A_{II}\omega)\bm c_{II}
+\bm y_{p,II}
\right],
\qquad
\omega=b_c-a_c .
\label{eq:phi_from_state}
\end{equation}
The equality \(\bm y_{II}(b_c)=\bm y_{III}(b_c)\) follows from state continuity. Thus \(\varphi(b_c)\) is not fitted independently; it is extracted from the normal Timoshenko matching solution.

\subsection{Timoshenko rotation-induced shear}

The reduced mixed-mode correction uses the Timoshenko section rotation, not the slope of the deflection, to estimate the basal tangential displacement induced by bending:
\begin{equation}
u_{\rm bend}=-\frac{h}{2}\varphi(b_c).
\label{eq:timo_ubend_app}
\end{equation}
With \(k_t=\tau_p/u_p\), this gives
\begin{equation}
\tau_{\rm bend}^{T}
=-k_t\frac{h}{2}\varphi(b_c).
\label{eq:timo_taubend_app}
\end{equation}
The total shear stress entering the reduced mixed-mode surface is
\begin{equation}
\tau_{\rm tip}
=
\tau_g\left(1+\frac{b_c}{\Lambda}\right)
+
\tau_{\rm bend}^{T},
\qquad
\Lambda=\sqrt{\frac{E'h}{k_t}}.
\label{eq:timo_tautip_app}
\end{equation}
The first contribution is the axial shear-transfer amplification of the gravitational shear load; the second is generated by Timoshenko slab rotation.

In the sharp-front limit, the Timoshenko matching gives the useful reference expressions
\[
\sigma_{\rm tip}
=
\sigma_g+
(\sigma_g-\sigma_r)
\left[
2p_T\frac{a_c}{\Lambda_b}
+
\left(\frac{a_c}{\Lambda_b}\right)^2
\right],
\]
and
\begin{equation}
|\varphi^{(0)}(a_c)|
=
2\,\frac{\sigma_g-\sigma_r}{k_n}
\frac{a_c}{\Lambda_b^2}
\left(1+p_T\frac{a_c}{\Lambda_b}\right),
\label{eq:timo_phi_sharp_app}
\end{equation}
so that
\begin{equation}
\tau_{\rm bend}^{(0,T)}
=
\pm(\sigma_g-\sigma_r)
\frac{k_t}{k_n}
\frac{h a_c}{\Lambda_b^2}
\left(1+p_T\frac{a_c}{\Lambda_b}\right).
\label{eq:timo_taubend_sharp_app}
\end{equation}
The sign is fixed by the slope-angle and cutting convention. Equation~\eqref{eq:timo_taubend_sharp_app} is the compact Timoshenko replacement for the bending-induced shear correction.

\subsection{Closed reduced mixed-mode system}

The finite-softening reduced mixed-mode problem is obtained by solving the Timoshenko normal matching conditions together with the peak condition
\begin{equation}
\left[
\frac{
\tau_g(1+b_c/\Lambda)-k_t h\varphi(b_c)/2
}{\tau_p}
\right]^2
+
\left(
\frac{\sigma_{\rm tip}}{\sigma_p}
\right)^2
=1.
\label{eq:timo_mixed_closed_app}
\end{equation}
Once \(\sigma_{\rm tip}\), \(a_c\), and \(b_c\) are known, the model provides both the fully collapsed length and the total affected length. The construction recovers the pure compressive Timoshenko anticrack when the shear contribution vanishes. The solution is analytical in the sense that the fields in each zone are exact matrix-exponential solutions of constant-coefficient Timoshenko equations; the crack lengths and integration constants are then obtained from a finite nonlinear algebraic system. It is not, in general, a single closed scalar formula for \(a_c(\delta_n)\).

\section{Fully coupled Timoshenko cohesive model}
\label{app:fully_coupled_timoshenko}

The reduced mixed-mode correction above combines a normal Timoshenko anticrack field with a separate shear-transfer estimate. Here we give the corresponding fully coupled formulation, in which the axial displacement, transverse deflection, section rotation, normal support, and tangential support are solved in a single boundary-value problem. The sharp-crack fully coupled solution used as a reference in the main text is not a separate theory; it is the brittle limit of the cohesive formulation obtained when the process-zone length tends to zero. This appendix therefore presents the finite-softening model first and then recovers the sharp-crack solution as the limit \(\delta_t/u_p,\delta_n/w_p\to0\), for which \(b_c-a_c\to0\) and \(a_c=b_c\).

\subsection{Coupled kinematics and weak-layer tractions}

The coupled state vector is
\[
\bm z=
\begin{bmatrix}
u & u' & w & w' & \varphi & \varphi'
\end{bmatrix}^{\mathsf T},
\]
where \(u\) is the slope-parallel displacement of the slab reference axis, \(w\) is the slope-normal deflection, and \(\varphi\) is the Timoshenko section rotation. The slab resultants are
\[
N=A_{11}u'+B_{11}\varphi',
\qquad
M=B_{11}u'+D_{11}\varphi',
\qquad
V=K_s(w'+\varphi).
\]
For a homogeneous slab, \(B_{11}=0\), \(A_{11}=E'h\), and \(D_{11}=\mathcal B\).

In the intact weak layer, the normal and tangential tractions are evaluated from the same coupled kinematics:
\begin{equation}
\sigma=k_n w,
\label{eq:fully_sigma_app}
\end{equation}
\begin{equation}
\tau=k_t\left(\frac{h_w}{2}w'-u-\frac{h}{2}\varphi\right),
\label{eq:fully_tau_app}
\end{equation}
where \(h_w\) is the weak-layer thickness entering the shear kinematics. Equation~\eqref{eq:fully_tau_app} is the main difference between the fully coupled model and the reduced correction: shear is not added after solving the normal problem; it is part of the coupled boundary-value problem itself.

\subsection{Mixed-mode cohesive law}

Because the intact-side traction vector is
\[
\bm t=
\begin{bmatrix}
\sigma\\ \tau
\end{bmatrix}
=
\begin{bmatrix}
k_n w\\
k_t\left(h_w w'/2-u-h\varphi/2\right)
\end{bmatrix},
\]
a process zone must prescribe how the normal and tangential components degrade together after the mixed-mode peak surface is reached. The scalar normal softening branch used in the reduced anticrack analogue is therefore not sufficient for the fully coupled problem.

We use an associated linear softening rule in normalized traction space. Define
\[
\widehat{\bm t}
=
\begin{bmatrix}
\widehat\sigma\\ \widehat\tau
\end{bmatrix}
=
\begin{bmatrix}
\sigma/\sigma_p\\ \tau/\tau_p
\end{bmatrix},
\qquad
F(\widehat{\bm t})=\widehat\sigma^2+\widehat\tau^2-1 .
\]
At the leading edge of the process zone, \(x=b_c\), the stress state lies on \(F=0\). We write the corresponding mixed-mode direction as
\[
\bm n=
\begin{bmatrix}
n_\sigma\\ n_\tau
\end{bmatrix}
=
\begin{bmatrix}
\sigma_{\rm tip}/\sigma_p\\
\tau_{\rm tip}/\tau_p
\end{bmatrix},
\qquad
n_\sigma^2+n_\tau^2=1 .
\]
This vector is the outward normal to the unit peak surface in normalized stress space. This normalization is important: the normal to the ellipse in the physical \((\sigma,\tau)\) plane is generally not radial and therefore does not, in general, pass through the origin. Linear degradation to \((\sigma,\tau)=(0,0)\) is therefore imposed along the normal in normalized stress space.

Let the tangential separation entering the fully coupled shear law be
\[
\Delta_t=\frac{h_w}{2}w'-u-\frac{h}{2}\varphi,
\]
so that the intact elastic law is
\[
\sigma=k_n w,\qquad \tau=k_t\Delta_t,
\qquad
w_p=\frac{\sigma_p}{k_n},\qquad
u_p=\frac{\tau_p}{k_t}.
\]
For a fixed mixed-mode direction \(\bm n\), introduce the normalized separation coordinate
\begin{equation}
s
=
n_\sigma\frac{w}{w_p}
+
n_\tau\frac{\Delta_t}{u_p}.
\label{eq:mixed_soft_s_app}
\end{equation}
At the onset of softening, \(s=1\). A linear softening law to zero traction can then be written as
\begin{equation}
\begin{bmatrix}
\sigma\\ \tau
\end{bmatrix}
=
\begin{bmatrix}
\sigma_p n_\sigma\\
\tau_p n_\tau
\end{bmatrix}
\left[
1-\frac{s-1}{\bar\delta_m}
\right],
\qquad
1\le s\le 1+\bar\delta_m ,
\label{eq:mixed_soft_law_app}
\end{equation}
where \(\bar\delta_m\) is a dimensionless mixed-mode softening distance. At \(s=1+\bar\delta_m\), both stress components vanish. The pure-mode limits are recovered by requiring
\[
\bar\delta_m(1,0)=\frac{\delta_n}{w_p},
\qquad
\bar\delta_m(0,1)=\frac{\delta_t}{u_p},
\]
where \(\delta_t\) is the tangential softening distance. A simple interpolation is
\[
\bar\delta_m(\bm n)
=
n_\sigma^2\frac{\delta_n}{w_p}
+
n_\tau^2\frac{\delta_t}{u_p},
\]
or, if only one softening distance is available, one may set
\(\delta_t/u_p=\delta_n/w_p\).

\subsection{Three-segment finite-softening solution}

For a prescribed mixed-mode direction \(\bm n\), Eqs.~\eqref{eq:mixed_soft_s_app}--\eqref{eq:mixed_soft_law_app} are affine in the state variables. Thus the fully coupled finite-softening problem remains linear with constant coefficients inside each segment. The affected zone is decomposed as
\[
\text{segment I: }0\le x\le a_c,
\qquad
\text{segment II: }a_c\le x\le b_c,
\qquad
\text{segment III: }0\le \xi\le L_\infty ,
\]
where segment I is the fully collapsed zone, segment II is the process zone, and segment III is the intact supported region measured from the leading edge with \(\xi=x-b_c\). In each segment,
\begin{equation}
\bm z'=\bm K_j\bm z+\bm d_j,
\label{eq:fully_state_app}
\end{equation}
where \(\bm K_j\) contains the Timoshenko slab stiffnesses and the appropriate weak-layer tangent stiffnesses. Segment I has no cohesive support in the void-like anticrack analogue, segment III uses the intact stiffnesses \(k_n\) and \(k_t\), and segment II uses the tangent matrix implied by the mixed-mode softening law. The solution is assembled from exact homogeneous modes and exact particular solutions:
\[
\bm z_j(x)=\bm Z_{h,j}(x)\bm c_j+\bm z_{p,j}(x).
\]

The constants are obtained from force-free conditions at the crack mouth, continuity at the collapsed/process-zone interface, continuity at the leading edge, and far-field equilibrium in the intact segment. Defining
\[
\bm D\bm z=
\begin{bmatrix}u&w&\varphi\end{bmatrix}^{\mathsf T},
\qquad
\bm R\bm z=
\begin{bmatrix}N&M&V\end{bmatrix}^{\mathsf T},
\]
the mechanical matching conditions are
\[
\bm R\bm z_I(0)=\bm 0,
\]
\[
\bm D\bm z_I(a_c)=\bm D\bm z_{II}(a_c),
\qquad
\bm R\bm z_I(a_c)=\bm R\bm z_{II}(a_c),
\]
\[
\bm D\bm z_{II}(b_c)=\bm D\bm z_{III}(0),
\qquad
\bm R\bm z_{II}(b_c)=\bm R\bm z_{III}(0),
\]
\[
\bm R\bm z_{III}(L_\infty)=\bm 0.
\]
The softening-zone end conditions are
\[
s(a_c)=1+\bar\delta_m(\bm n),
\qquad
s(b_c)=1,
\]
and the leading-edge stress state satisfies \(F(\bm n)=0\). The unknowns are therefore the integration constants, the collapsed length \(a_c\), the leading-edge length \(b_c\), and the mixed-mode direction \(\bm n\). The resulting formulation is exact within each linear segment, but the final crack lengths are obtained from a finite nonlinear algebraic system. It is therefore a semi-analytical cohesive-zone solution rather than a compact scalar formula for \(a_c\).

\subsection{Sharp-crack limit}

The fully coupled sharp-crack solution follows from the limit
\[
\frac{\delta_t}{u_p}\to0,
\qquad
\frac{\delta_n}{w_p}\to0,
\qquad
b_c-a_c\to0 .
\]
The process-zone segment then disappears and the problem reduces to two segments: an unsupported collapsed segment and an intact supported segment,
\[
\text{segment I: }0\le x\le a,
\qquad
\text{segment II: }0\le \xi\le L_\infty ,
\]
where segment II is measured from the sharp front and \(a=a_c=b_c\). The same state equation \eqref{eq:fully_state_app} is used in each segment, with no cohesive support in segment I and intact weak-layer stiffnesses in segment II. The matching conditions reduce to
\[
\bm R\bm z_I(0)=\bm 0,
\]
\[
\bm D\bm z_I(a)=\bm D\bm z_{II}(0),
\qquad
\bm R\bm z_I(a)=\bm R\bm z_{II}(0),
\]
\[
\bm R\bm z_{II}(L_\infty)=\bm 0.
\]
The crack-tip stresses are evaluated on the intact side:
\[
\sigma_{\rm tip}=k_n w_{\rm tip},
\]
\[
\tau_{\rm tip}=k_t\left(\frac{h_w}{2}w'_{\rm tip}-u_{\rm tip}-\frac{h}{2}\varphi_{\rm tip}\right).
\]
The brittle propagation condition is
\begin{equation}
\left(\frac{\tau_{\rm tip}}{\tau_p}\right)^2
+
\left(\frac{\sigma_{\rm tip}}{\sigma_p}\right)^2
=1.
\label{eq:fully_peak_app}
\end{equation}
The critical length is the first positive root of Eq.~\eqref{eq:fully_peak_app} as a function of \(a\). This sharp-crack limit is useful because it isolates the effect of full Timoshenko coupling from the additional effect of finite softening, and because it provides the closest counterpart to WEAC-like sharp-crack formulations.

\section{Laboratory experiments}
The shear tests were performed using a linear friction tester (LFT), based on the apparatus described by \citet{Theile2009}. The LFT consists of two high-power, high-precision computer-controlled linear drives capable of imposing prescribed two-dimensional displacement paths. In the experiments considered here, the vertical drive was deactivated, so that the upper sample holder rested freely on the snow sample under its own weight. Because the upper holder was relatively heavy, its weight was partly compensated by a counterweight to avoid excessive initial compaction or crushing of the samples. The snow sample was frozen between the lower and upper plates: the plates were first warmed with a blow dryer, after which the upper plate was left resting on the sample for several minutes until refreezing occurred. The horizontal drive then displaced the upper plate at a prescribed constant velocity, imposing direct shear under displacement control. Force and displacement were recorded at sampling rates between \(1\) and \(10~\mathrm{Hz}\). The three force components were measured with a Kistler Model~9254 piezoelectric force plate placed under the lower sample holder, with a force resolution of \(0.1~\mathrm{N}\), while displacement was measured using a Renishaw linear encoder with a resolution of \(1~\mu\mathrm{m}\).

\section{Numerical modeling}

We simulated Propagation Saw Tests (PSTs) using two complementary Material Point Method formulations to determine both the critical crack length \(a_c\), defined by the saw position, and the total affected length \(b_c\), inferred from the onset and development of plastic deformation in the weak layer. We used a three-dimensional Material Point Method (MPM) framework to simulate anticrack propagation, for which weak-layer collapse, volumetric plastic deformation, and slab bending must be resolved explicitly. In addition, we used the Depth-Averaged Material Point Method (DAMPM), a reduced depth-integrated formulation, to simulate shear-failure propagation governed by a strain-softening weak-layer interface.

Both formulations use an elastic slab in order to isolate weak-layer-controlled propagation and remove the possible influence of slab fracture and associated crack arrest. However, the weak-layer representation differs between the two approaches to reflect the distinct failure mechanisms considered. The principal material parameters and constitutive assumptions are summarized in Table~\ref{tab:numerical_parameters}. Further details on the numerical algorithms, implementations, and validation of the three-dimensional MPM and DAMPM formulations can be found in \citet{gaume_nc_2018,trottet2022} and \citet{guillet2023,meloche2025modeling}, respectively.

\subsection{Three-dimensional Material Point Method: anticrack propagation}

For the anticrack simulations, we used the elastoplastic MPM framework described by \citet{trottet2022}. In this formulation, the weak layer is represented as a compressible elastoplastic material, while the slab is modeled as an elastic continuum. This depth-resolved representation allows weak-layer collapse, slab bending, and the resulting redistribution of stresses to be captured explicitly.

The weak layer is described using an associative cohesive Cam--Clay yield surface,
\begin{equation}
    y_{wl}(p,q) = q^2(1+2\beta) + M^2(p+\beta p_0)(p-p_0),
\end{equation}
where \(p\) is the pressure and \(q\) is the von Mises equivalent stress. The parameter \(p_0\) denotes the preconsolidation pressure, \(\beta\) is the cohesion parameter, and \(M\) is the slope of the critical-state line in the absence of cohesion.

To capture strain softening associated with weak-layer collapse, the preconsolidation pressure was allowed to evolve accordingly to
\begin{equation}
    p_0^{wl} = K \sinh \!\left(\xi \max(\eta,0) \right),
\end{equation}
where \(K\) is the bulk modulus, $\xi$ is a hardening parameter, and
\begin{equation}
    \dot\eta = \gamma |\dot\epsilon_V^p|
\end{equation}
is the anticrack plastic strain, with \(\epsilon_V^p\) denoting the plastic volumetric strain and \(\gamma\) a softening parameter. This constitutive formulation enables progressive weak-layer collapse and strain softening, leading to the propagation of an anticrack beneath the elastic slab. Further details regarding the constitutive law and its numerical implementation are provided by \citet{trottet2022}.

\subsection{Depth-Averaged Material Point Method: shear failure propagation}

For the shear-failure simulations, we used DAMPM, a depth-integrated Material Point Method formulation in which the slab motion is represented through depth-averaged quantities. DAMPM is derived from the same continuum-mechanics framework as full MPM but relies on a reduced kinematic description, making it particularly suitable for efficiently simulating shear-dominated propagation over extended domains. The formulation and its application to snow-avalanche release and propagation are described in \citet{guillet2023,meloche2025modeling}.

In the DAMPM simulations, the weak layer is represented as a quasi-brittle interface governed by a strain-softening shear law. The interface shear stress increases linearly with tangential displacement until reaching the peak shear strength \(\tau_p\) at the critical displacement
\begin{equation}
    u_c = \frac{\tau_p}{k_t},
\end{equation}
where \(k_t\) is the tangential stiffness of the weak layer.

Once the tangential displacement exceeds this critical value, i.e., \(u_x > u_c\), the shear stress \(\tau_{xz}\) decreases linearly from the peak strength \(\tau_p\) towards the residual strength \(\tau_r\). The residual strength is reached when the displacement attains the residual displacement
\begin{equation}
    u_r = u_c + \delta,
\end{equation}
where \(\delta\) is the softening displacement. This interface law allows the progressive development of a shear fracture process zone and the propagation of a shear crack along the weak layer.

\begin{table}[t]
    \centering
    \caption{Material and simulation parameters used in the three-dimensional MPM anticrack simulations and the DAMPM shear-failure simulations. A dash indicates that the parameter is not used in the corresponding formulation.}
    \label{tab:numerical_parameters}
    \begin{tabular}{lcc}
        \hline
        Parameter & 3D MPM (anticrack) & DAMPM (shear) \\
        \hline
        \multicolumn{3}{l}{\textit{Slab properties}} \\
        Slab thickness, \(h\) & \(0.5~\mathrm{m}\) & \(0.5~\mathrm{m}\) \\
        Slab density, \(\rho\) & \(250~\mathrm{kg\,m^{-3}}\) & \(250~\mathrm{kg\,m^{-3}}\) \\
        Slab Young's modulus, \(E\) & \(4 \times 10^{6}~\mathrm{Pa}\) & \(4 \times 10^{6}~\mathrm{Pa}\) \\
        Poisson's ratio, \(\nu\) & \(0.3\) & \(0.3\) \\
        \hline
        \multicolumn{3}{l}{\textit{Weak-layer parameters}} \\
        Peak shear strength, \(\tau_p\) & -- & \(1800~\mathrm{Pa}\) \\
        Tangential stiffness, \(k_t\) & -- & \(5 \times 10^{6}~\mathrm{Pa\,m^{-1}}\) \\
        Weak-layer Young's modulus, \(E_{\rm wl}\) & \(1 \times 10^{6}~\mathrm{Pa}\) & -- \\
        Preconsolidation pressure, \(p_0\) & \(13 \times 10^{3}~\mathrm{Pa}\) & -- \\
        Critical-state-line slope, \(M\) & \(0.8\) & -- \\
        Cohesion parameter, \(\beta\) & \(0.2\) & -- \\
        Friction angle, \(\phi\) & \(27^\circ\) & \(27^\circ\) \\
        \hline
        \multicolumn{3}{l}{\textit{Simulation configuration}} \\
        Slope angle, \(\theta\) & \(0^\circ\) & \(35^\circ\) \\
        \hline
    \end{tabular}
\end{table}

\newpage
\section*{Notation}

\renewcommand{\arraystretch}{1.12}
\begin{longtable}{p{0.18\textwidth} p{0.74\textwidth}}
\hline
Symbol & Meaning \\
\hline
\endfirsthead

\hline
Symbol & Meaning \\
\hline
\endhead

\multicolumn{2}{l}{\textit{Coordinates, kinematics, and common slab properties}} \\
\hline

\(x\) & Slope-parallel coordinate. \\
\(z\) & Vertical coordinate, positive upward, with \(z=0\) at the slab free surface and \(z=-h\) at the slab--weak-layer interface. \\
\(u(x)\) & Slope-parallel weak-layer displacement in the shear formulation. \\
\(w(x)\) & Slope-normal slab deflection in the anticrack extension. \\
\(h\) & Slab thickness. \\
\(\rho\) & Slab density. \\
\(E\) & Young's modulus of the elastic slab in the numerical simulations. \\
\(E'\) & Plane-strain elastic modulus of the slab in the analytical formulation. \\
\(\nu\) & Poisson's ratio of the slab. \\
\(E'h\) & Axial stiffness of the slab per unit width. \\
\(\sigma_{xx}\) & Slope-parallel normal stress in the slab. \\
\(\theta\) & Slope angle. \\
\(\tau_g\) & Gravitational shear stress acting on the slab. \\
\(\sigma_g\) & Gravitational normal stress acting on the slab. \\

\\[-0.3em]
\hline
\multicolumn{2}{l}{\textit{Shear failure model}} \\
\hline

\(\tau(x)\) & Shear stress transmitted by the weak layer. \\
\(\tau_p\) & Peak shear strength of the weak layer. \\
\(\tau_r\) & Residual shear strength of the weak layer. \\
\(u_p\) & Displacement at peak shear strength. \\
\(u_r\) & Displacement at residual shear strength. \\
\(\delta=u_r-u_p\) & Softening displacement in shear. \\
\(k_t\) & Pre-peak tangential stiffness of the weak layer, \(k_t=\tau_p/u_p\). \\
\(\Lambda\) & Elastic characteristic length of the slab--weak-layer system, \(\Lambda=\sqrt{E'h\,u_p/\tau_p}\). \\
\(\ell\) & Shear-softening length, \(\ell=\sqrt{E'h\,\delta/(\tau_p-\tau_r)}\). \\
\(\alpha\) & Dimensionless softening-zone length, \(\alpha=(b_c-a_c)/\ell\). \\
\(a_c\) & Critical crack length, i.e.\ half-length of the fully softened residual zone at propagation. \\
\(b_c\) & Outer half-length of the affected zone at propagation, including both the fully softened and softening regions. \\
\(\omega=b_c-a_c\) & Fracture-process-zone length, or softening-zone length. \\
\(a_{c0}\) & Brittle critical crack length obtained in the limit \(\delta\rightarrow0\). \\
\(C_a\) & Dimensionless coefficient entering the exact square-root expression for \(a_c\) and its linear small-softening limit. \\
\(C_b\) & Dimensionless coefficient entering the linear small-softening approximation of \(b_c\). \\
\(u_I,u_{II},u_{III}\) & Displacement solutions in the residual, softening, and intact zones, respectively. \\
\(\tau_I,\tau_{II},\tau_{III}\) & Shear-stress solutions in the residual, softening, and intact zones, respectively. \\
\(u_0\) & Displacement at \(x=0\) in the residual-zone solution. \\
\(u_c\) & Constant particular solution in the shear softening zone. \\
\(A,B\) & Integration constants in the shear softening-zone solution. \\
\(\ddot{u}_{\mathrm{res}}\) & Average slope-parallel acceleration in the residual zone used for the local dynamic correction in the DAMPM comparison. \\

\\[-0.3em]
\hline
\multicolumn{2}{l}{\textit{Energy-based sharp-crack interpretation}} \\
\hline

\(G(a)\) & Energy-release rate of the corresponding sharp-crack slab--weak-layer system. \\
\(G_{\mathrm{soft}}\) & Post-peak cohesive degradation energy measured above the residual stress level, \(G_{\mathrm{soft}}=\frac12(\tau_p-\tau_r)\delta\). \\
\(G_{IIc}^{\mathrm{intr}}\) & Intrinsic bond-breaking energy of the weak-layer interface. \\
\(G_{IIc}^{\mathrm{LEFM}}\) & Effective mode-II fracture energy used in the equivalent sharp-crack LEFM description. \\
\(W_{\mathrm{fric}}\) & Residual frictional work associated with sliding of the already failed interface. \\
\(a_{\mathrm{LEFM}}\) & Equivalent sharp-crack critical length obtained from the LEFM energy balance. \\
\(\Pi(a)\) & Total potential energy per unit width of the sharp-crack slab--weak-layer system, defined up to an \(a\)-independent reference energy. \\

\\[-0.3em]
\hline
\multicolumn{2}{l}{\textit{Mixed-mode anticrack energy and equivalent toughness}} \\
\hline

\(G_I\) & Mode-I energy-release-rate measure associated with weak-layer compression in the WEAC-like sharp-crack interpretation, \(G_I=\sigma_{\rm tip}^2/(2k_n)\). \\
\(G_{II}\) & Mode-II energy-release-rate measure associated with weak-layer shear in the WEAC-like sharp-crack interpretation, \(G_{II}=\tau_{\rm tip}^2/(2k_t)\). \\
\(G_{Ic}\) & Equivalent mode-I fracture toughness derived from the normal weak-layer constitutive law. \\
\(G_{IIc}\) & Equivalent mode-II fracture toughness derived from the tangential weak-layer constitutive law. \\
\(G_{Ic}^{(0)}\) & Zero-softening equivalent mode-I toughness, \(G_{Ic}^{(0)}=\sigma_p^2/(2k_n)=\frac12\sigma_p w_p\). \\
\(G_{IIc}^{(0)}\) & Zero-softening equivalent mode-II toughness, \(G_{IIc}^{(0)}=\tau_p^2/(2k_t)=\frac12\tau_p u_p\). \\
\(G_{Ic}^{\rm eff}\) & Effective sharp-crack mode-I toughness obtained by integrating the normal cohesive work above residual support. \\
\(G_{IIc}^{\rm eff}\) & Effective sharp-crack mode-II toughness obtained by integrating the tangential cohesive work above residual resistance. \\

\\[-0.3em]
\hline
\multicolumn{2}{l}{\textit{Timoshenko anticrack, bending, and mixed-mode extension}} \\
\hline

\(\mathcal{B}\) & Slab bending stiffness in the Timoshenko anticrack extension, \(\mathcal B=E'h^3/12\). \\
\(K_s\) & Timoshenko transverse-shear rigidity of the slab, \(K_s=\kappa Gh\). \\
\(\varphi(x)\) & Timoshenko cross-section rotation of the slab. \\
\(M,V\) & Slab bending moment and transverse shear force in the Timoshenko anticrack formulation. \\
\(\sigma(w)\) & Slope-normal weak-layer support law in the anticrack extension. \\
\(\sigma_p\) & Peak compressive weak-layer strength in the anticrack extension. \\
\(\sigma_r\) & Residual compressive weak-layer support in the anticrack extension. \\
\(E_{\rm wl}\) & Effective normal or Young's modulus of the weak layer, used to define the normal foundation stiffness over the weak-layer thickness. \\
\(G_{\rm wl}\) & Effective shear modulus of the weak layer, used to define the tangential foundation stiffness over the weak-layer thickness. \\
\(w_p\) & Deflection at peak compressive support. \\
\(w_r\) & Deflection at residual compressive support. \\
\(\delta_n=w_r-w_p\) & Softening displacement in the normal/compressive anticrack formulation. \\
\(k_n\) & Pre-peak compressive stiffness of the weak layer, \(k_n=\sigma_p/w_p\). \\
\(k_s\) & Tangent softening stiffness in compression, \(k_s=(\sigma_p-\sigma_r)/\delta_n\). \\
\(k_s^{\rm tip}\) & Tangent softening stiffness in the mixed-mode anticrack formulation, \(k_s^{\rm tip}=\sigma_{\rm tip}/\delta_n\). \\
\(\Lambda_b\) & Bending--foundation characteristic length, \(\Lambda_b=(4\mathcal{B}/k_n)^{1/4}\). \\
\(\eta_B\) & Timoshenko shear-deformation parameter, \(\eta_B=\mathcal B/(K_s\Lambda_b^2)\). \\
\(p_T\) & Timoshenko correction factor, \(p_T=\sqrt{1+\eta_B}\). \\
\(\ell_b\) & Bending--softening length, \(\ell_b=(\mathcal{B}\delta_n/(\sigma_p-\sigma_r))^{1/4}\). \\
\(\ell_b^{\rm tip}\) & Mixed-mode bending--softening length, \(\ell_b^{\rm tip}=(\mathcal{B}\delta_n/\sigma_{\rm tip})^{1/4}\). \\
\(\ell_s\) & Timoshenko shear-deformation softening length, \(\ell_s=(K_s/k_s)^{1/2}\). \\
\(\ell_s^{\rm tip}\) & Mixed-mode Timoshenko shear-deformation softening length, \(\ell_s^{\rm tip}=(K_s\delta_n/\sigma_{\rm tip})^{1/2}\). \\
\(\bm y\) & Timoshenko normal anticrack state vector, \(\bm y=[w,\varphi,M,V]^{\mathsf T}\). \\
\(\tau_{\rm bend}^{T}\) & Shear stress induced by Timoshenko slab rotation in the reduced mixed-mode anticrack extension. \\
\(\tau_{\rm tip}\) & Shear stress at the leading edge of the affected zone in the mixed-mode extension. \\
\(\sigma_{\rm tip}\) & Normal stress at the leading edge of the affected zone in the mixed-mode extension. \\
\(w_{\rm tip}\) & Normal displacement at the leading edge of the affected zone in the mixed-mode extension. \\
\(C_a^{(b,T)}, C_b^{(b,T)}\) & Dimensionless coefficients in the approximate finite-softening Timoshenko anticrack expressions for \(a_c\) and \(b_c\). \\
\(\bm z\) & Fully coupled Timoshenko state vector, \(\bm z=[u,u',w,w',\varphi,\varphi']^{\mathsf T}\). \\
\(h_w\) & Weak-layer thickness entering the fully coupled tangential kinematics. \\
\(\Delta_t\) & Tangential separation entering the fully coupled cohesive law, \(\Delta_t=h_w w'/2-u-h\varphi/2\). \\
\(\bm n=(n_\sigma,n_\tau)^{\mathsf T}\) & Mixed-mode direction on the normalized peak surface, with \(n_\sigma^2+n_\tau^2=1\). \\
\(\bar\delta_m\) & Dimensionless mixed-mode softening distance in the associated normalized cohesive law. \\
\(\delta_t\) & Tangential softening distance used to define the pure shear limit of \(\bar\delta_m\). \\
\(q_{\rm surf}\) & Additional uniform surface load applied on top of the slab. \\
\(c_{\rm ecc}\) & Fraction of the additional surface load transmitted through the top-surface eccentric load path. \\

\\[-0.3em]
\hline
\multicolumn{2}{l}{\textit{Three-dimensional MPM weak-layer constitutive parameters}} \\
\hline

\(p\) & Pressure in the weak layer. \\
\(q\) & von Mises equivalent stress in the weak layer. \\
\(y_{wl}(p,q)\) & Cohesive Cam--Clay yield function of the weak layer. \\
\(K\) & Bulk modulus of the weak layer in the MPM formulation. \\
\(p_0^{wl}\) & Evolving preconsolidation pressure of the weak layer. \\
\(M\) & Slope of the critical-state line in the absence of cohesion. \\
\(\beta\) & Cohesion parameter in the cohesive Cam--Clay yield surface. \\
\(\phi\) & Friction angle of the weak layer. \\
\(\eta\) & Internal accumulated anticrack/collapse strain variable controlling the evolution of \(p_0^{wl}\). \\
\(\epsilon_V^p\) & Plastic volumetric strain of the weak layer. \\
\(\dot{\epsilon}_V^p\) & Rate of plastic volumetric strain of the weak layer. \\
\(\gamma\) & Scaling parameter relating the evolution of \(\eta\) to the plastic volumetric strain rate, \(\dot{\eta}=\gamma |\dot{\epsilon}_V^p|\). \\
\(\xi\) & Hardening parameter controlling the sensitivity of the preconsolidation pressure to \(\eta\). \\

\\[-0.3em]
\hline
\multicolumn{2}{l}{\textit{Auxiliary quantities for the small-softening expansion}} \\
\hline

\(T\) & Auxiliary notation in Appendix~\ref{app:small_delta}, \(T=\tau_p-\tau_g\). \\
\(R\) & Auxiliary notation in Appendix~\ref{app:small_delta}, \(R=\tau_g-\tau_r\). \\
\(s\) & Small parameter used in Appendix~\ref{app:small_delta}, \(s=\ell/\Lambda\). \\
\(m\) & Auxiliary ratio used in Appendix~\ref{app:small_delta}, \(m=(\tau_p-\tau_g)/(\tau_g-\tau_r)\). \\

\hline
\end{longtable}

\printbibliography[heading=bibintoc]

@String { AoG = {Ann. Glaciol.} }

@String { CR  = {Cold Reg. Sci. Technol.} }

@String { JGR = {J. Geophys. Res.} }

@String { JoG = {J. Glaciol.} }

@Article{Chiaia2008,
  Title                    = {Triggering of dry snow slab avalanches: stress versus fracture mechanical approach},
  Author                   = {Chiaia, B.M and Cornetti, P and Frigo, B},
  Journal                  = CR,
  Year                     = {2008},
  Pages                    = {170-178},
  Volume                   = {53}
}

@Article{Gaume2013a,
  Title                    = {Influence of weak-layer heterogeneity on snow slab avalanche release: Application to the evaluation of avalanche release depths.},
  Author                   = {Gaume, J. and Chambon, G. and Eckert, N. and Naaim, M.},
  Journal                  = JoG,
  Year                     = {2013},
  Pages                    = {423-437},
  Volume                   = {59(215)}
}

@Article{Gaume2018,
  Title                    = {Dynamic anticrack propagation in snow},
  Author                   = {Gaume, J. and Gast, T. and Teran, J. and van Herwijnen, A. and Jiang, C.},
  Journal                  = {Nature Communications},
  Year                     = {2018},
  Number                   = {1},
  Pages                    = {3047},
  Volume                   = {9}
}

@Article{Gaume2015,
  Title                    = {Modeling of crack propagation in weak snowpack layers using the discrete element method},
  Author                   = {Gaume, J. and van Herwijnen, A. and Chambon, G. and Birkeland, K. and Schweizer, J.},
  Journal                  = {The Cryosphere},
  Year                     = {2015},
  Pages                    = {1915-1932},
  Volume                   = {9}
}

@Article{Gaume2017,
  Title                    = {Snow fracture in relation to slab avalanche release: critical state for the onset of crack propagation},
  Author                   = {Gaume, J. and van Herwijnen, A. and Chambon, G. and Wever, N. and Schweizer, J.},
  Journal                  = {The Cryosphere},
  Year                     = {2017},
  Pages                    = {217-228},
  Volume                   = {11},

  Doi                      = {doi:10.5194/tc-2016-64}
}

@Article{Gaume2014,
  Title                    = {Evaluation of slope stability with respect to snowpack spatial variability},
  Author                   = {Gaume, J. and Schweizer, J. and van Herwijnen, A. and Chambon, G. and Reuter, B. and Eckert, N. and Naaim, M.},
  Journal                  = JGR,
  Year                     = {2014},
  Number                   = {9},
  Pages                    = {1783-1789},
  Volume                   = {119},

  Doi                      = {10.1002/2014JF003193}
}

@PhdThesis{Heierli2008_,
  Title                    = {Anticrack model for slab avalanche release},
  Author                   = {Heierli, Joachim},
  School                   = {Karlsruhe, Univ., Diss.},
  Year                     = {2008}
}

@Article{Heierli2008,
  Title                    = {Anticrack nucleation as triggering mechanism for snow slab avalanches},
  Author                   = {Heierli, J and Gumbsch, P and Zaiser, M.},
  Journal                  = {Science},
  Year                     = {2008},
  Pages                    = {240--243},
  Volume                   = {321}
}

@Article{Mahajan2010,
  Title                    = {Numerical simulation of failure in a layered thin snowpack under skier load},
  Author                   = {Mahajan, P. and Kalakuntla, R. and Chandel, C.},
  Journal                  = AoG,
  Year                     = {2010},
  Pages                    = {169-175},
  Volume                   = {51(54)}
}

@Article{McClung1979,
  Title                    = {Shear fracture precipitated by strain softening as a mechanism of dry slab avalanche release},
  Author                   = {McClung, D.M.},
  Journal                  = JGR,
  Year                     = {1979},
  Pages                    = {3519-3526},
  Volume                   = {84(B7)}
}

@Article{McClung1977,
  Title                    = {Direct simple shear tests on snow and their relation to slab avalanche formation},
  Author                   = {McClung, D.M.},
  Journal                  = JoG,
  Year                     = {1977},
  Pages                    = {101-109},
  Volume                   = {19(81)}
}

@Article{Monti2016,
  Title                    = {A simplified approach to assess the skier-induced stress within a multi-layered snowpack},
  Author                   = {Monti, F. and Gaume, J. and Van Herwijnen, A. and Schweizer, J.},
  Journal                  = {Nat. Hazard Earth Syst. Sci.},
  Year                     = {2016},
  Volume                   = {16}
}

@Article{Palmer1973,
  Title                    = {The growth of slip surfaces in the progressive failure of over-consolidated clay},
  Author                   = {Palmer, A.C and Rice, J.R},
  Journal                  = {Proc. R .Soc. London},
  Year                     = {1973},
  Pages                    = {527-548},
  Volume                   = {332}
}

@Article{Schweizer1998,
  Title                    = {Laboratory experiments on shear failure of snow},
  Author                   = {Schweizer, J},
  Journal                  = AoG,
  Year                     = {1998},
  Pages                    = {97-102},
  Volume                   = {26}
}

@PREAMBLE{ {\providecommand{\noopsort}[1]{}} }

@ARTICLE{gaume_nc_2018,
author={Gaume, J. and Gast, T. and Teran, J. and van Herwijnen, A. and Jiang, C.},
title={Dynamic anticrack propagation in snow},
journal={Nature Communications},
year={2018},
volume={9},
number={1},
doi={10.1038/s41467-018-05181-w},
art_number={3047},
url={https://www.scopus.com/inward/record.uri?eid=2-s2.0-85051092260&doi=10.1038\%2fs41467-018-05181-w&partnerID=40&md5=f3537aae0415c9064b431a674069f96e},
document_type={Article},
source={Scopus},
}

@article{trottet_np_2021,
author = {Trottet, Bertil and Simenhois, Ron and Bobillier, Grégoire and van Herwijnen, Alec and Jiang, Chenfanfu and Gaume, Johan},
year = {2021},
pages = {preprint},
title = {Transition from sub-Rayleigh anticrack to supershear crack propagation in snow avalanches},
journal = {Nature Physics},
doi = {https://doi.org/10.21203/rs.3.rs-963978/v1}
}

@article{schottner2025testing,
  title={Testing the strength of buried surface hoar weak layers under combined compression and shear loading},
  author={Sch{\"o}ttner, Jakob and Piecuch, Marcel and Walet, Melin and Rosendahl, Philipp and Adam, Valentin and Wei{\ss}graeber, Philipp and Schweizer, J{\"u}rg and van Herwijnen, Alec},
  journal={Authorea Preprints},
  year={2025},
  publisher={Authorea}
}

@article{adam2024fracture,
  title={Fracture toughness of mixed-mode anticracks in highly porous materials},
  author={Adam, Valentin and Bergfeld, Bastian and Wei{\ss}graeber, Philipp and van Herwijnen, Alec and Rosendahl, Philipp L},
  journal={Nature Communications},
  volume={15},
  number={1},
  pages={7379},
  year={2024},
  publisher={Nature Publishing Group UK London}
}

@article{weissgraeber2023closed,
  title={A closed-form model for layered snow slabs},
  author={Weissgraeber, Philipp and Rosendahl, Philipp L},
  journal={The Cryosphere},
  volume={17},
  number={4},
  pages={1475--1496},
  year={2023},
  publisher={Copernicus Publications G{\"o}ttingen, Germany}
}

@article{rosendahl2020modeling1,
  title={Modeling snow slab avalanches caused by weak-layer failure--Part 1: Slabs on compliant and collapsible weak layers},
  author={Rosendahl, Philipp L and Weissgraeber, Philipp},
  journal={The Cryosphere},
  volume={14},
  number={1},
  pages={115--130},
  year={2020},
  publisher={Copernicus Publications G{\"o}ttingen, Germany}
}

@article{rosendahl2020modeling2,
  title={Modeling snow slab avalanches caused by weak-layer failure--Part 2: Coupled mixed-mode criterion for skier-triggered anticracks},
  author={Rosendahl, Philipp L and Weissgraeber, Philipp},
  journal={The Cryosphere},
  volume={14},
  number={1},
  pages={131--145},
  year={2020},
  publisher={Copernicus Publications G{\"o}ttingen, Germany}
}

@article{bobillier2024numerical,
  title={Numerical investigation of crack propagation regimes in snow fracture experiments},
  author={Bobillier, Gr{\'e}goire and Bergfeld, Bastian and Dual, J{\"u}rg and Gaume, Johan and van Herwijnen, Alec and Schweizer, J{\"u}rg},
  journal={Granular matter},
  volume={26},
  number={3},
  pages={58},
  year={2024},
  publisher={Springer}
}

@article{siron2023theoretical,
  title={A theoretical framework for dynamic anticrack and supershear propagation in snow slab avalanches},
  author={Siron, Marin and Trottet, Bertil and Gaume, Johan},
  journal={Journal of the Mechanics and Physics of Solids},
  volume={181},
  pages={105428},
  year={2023},
  publisher={Elsevier}
}

@article{meloche2025modeling,
  title={Modeling crack arrest in snow slab avalanches—toward estimating avalanche release sizes},
  author={Meloche, Francis and Bobillier, Gr{\'e}goire and Guillet, Louis and Gauthier, Francis and Langlois, Alexandre and Gaume, Johan},
  journal={Journal of Geophysical Research: Earth Surface},
  volume={130},
  number={12},
  pages={e2025JF008470},
  year={2025},
  publisher={Wiley Online Library}
}

@unpublished{Volmer2017,
  author = {Volmer, J. C. and Anciaux, Guillaume and Peerlings, Ron and Gaume, Johan},
  title  = {Modelling Weak Layer Fracture Using Non-Associated Plasticity},
  note   = {MSc internship project report, {\'E}cole Polytechnique F{\'e}d{\'e}rale de Lausanne and Eindhoven University of Technology},
  year   = {2017},
  month  = jun
}

@phdthesis{reiweger2011failure,
  title={Failure of weak snow layers},
  author={Reiweger, Ingrid},
  year={2011},
  school={ETH Zurich}
}

@article{Puzrin2019,
  title={The mechanism of delayed release in earthquake-induced avalanches},
  author={Puzrin, Alexander M and Faug, Thierry and Einav, Itai},
  journal={Proceedings of the Royal Society A: Mathematical, Physical and Engineering Sciences},
  volume={475},
  number={2227},
  year={2019},
  publisher={The Royal Society}
}

@article{puzrin2005growth,
  title={The growth of shear bands in the catastrophic failure of soils},
  author={Puzrin, Alexander M and Germanovich, LN},
  journal={Proceedings of the Royal Society A: Mathematical, Physical and Engineering Sciences},
  volume={461},
  number={2056},
  pages={1199--1228},
  year={2005},
  publisher={The Royal Society London}
}

@article{Melin2026,
  author  = {Walet, Melin and Sch{\"o}ttner, Jakob and Kraus, Sirah and Adam, Valentin and Rosendahl, Philipp and Wei{\ss}graeber, Philipp and Rheinschmidt, Florian and Schweizer, J{\"u}rg and van Herwijnen, Alec},
  title   = {Field measurements of multiaxial strength of weak snow layers},
  journal = {Journal of Glaciology},
  year    = {2026},
  note    = {Submitted}
}

@book{Winkler1867,
  author    = {Winkler, Emil},
  title     = {Die Lehre von der Elasticitaet und Festigkeit mit besonderer R{\"u}cksicht auf ihre Anwendung in der Technik},
  publisher = {H. Dominicus},
  address   = {Prag},
  year      = {1867}
}

@article{Timoshenko1921,
  author  = {Timoshenko, S. P.},
  title   = {On the Correction for Shear of the Differential Equation for Transverse Vibrations of Prismatic Bars},
  journal = {The London, Edinburgh, and Dublin Philosophical Magazine and Journal of Science},
  series  = {6},
  volume  = {41},
  number  = {245},
  pages   = {744--746},
  year    = {1921},
  doi     = {10.1080/14786442108636264}
}

@article{guillet2023,
  title = {A {{Depth-Averaged Material Point Method}} for {{Shallow Landslides}}: {{Applications}} to {{Snow Slab Avalanche Release}}},
  shorttitle = {A {{Depth-Averaged Material Point Method}} for {{Shallow Landslides}}},
  author = {Guillet, Louis and Blatny, Lars and Trottet, Bertil and Steffen, Denis and Gaume, Johan},
  year = {2023},
  journal = {Journal of Geophysical Research: Earth Surface},
  volume = {128},
  number = {8},
  pages = {e2023JF007092},
  issn = {2169-9011},
  doi = {10.1029/2023JF007092},
  urldate = {2023-09-26},
  copyright = {{\copyright} 2023 The Authors.},
  langid = {english}
}

@article{trottet2022,
  title = {Transition from Sub-{{Rayleigh}} Anticrack to Supershear Crack Propagation in Snow Avalanches},
  author = {Trottet, Bertil and Simenhois, Ron and Bobillier, Gregoire and Bergfeld, Bastian and {van Herwijnen}, Alec and Jiang, Chenfanfu and Gaume, Johan},
  year = {2022},
  month = sep,
  journal = {Nature Physics},
  volume = {18},
  number = {9},
  pages = {1094--1098},
  publisher = {Nature Publishing Group},
  issn = {1745-2481},
  doi = {10.1038/s41567-022-01662-4},
  urldate = {2024-08-13},
  copyright = {2022 The Author(s)},
  langid = {english}
}

@article{Theile2009,
  author  = {Theile, T. and Szabo, D. and L{\"u}thi, A. and Rhyner, H. and Schneebeli, M.},
  title   = {Mechanics of the ski-snow contact},
  journal = {Tribology Letters},
  volume  = {36},
  number  = {3},
  pages   = {223--231},
  year    = {2009}
}

\end{document}